\documentclass[%
aps,
prd,
amssymb,
amsmath,
amsfonts,
eqsecnum,
showpacs,
nofootinbib,
floatfix,
twocolumn,
twoside,
a4paper,
superscriptaddress,
longbibliography
]{revtex4-1}

\usepackage[T3,T1]{fontenc}
\usepackage{mathrsfs} 
\usepackage{bm} 
\makeatletter
\@ifclassloaded{beamer}
  { 
    \typeout{UsePackages: Detected beamer}
    \usepackage{tgheros}
    
  }
  { 
    \typeout{UsePackages: Did not detect beamer}
    \ifx\asybeamer\undefined 
    \typeout{UsePackages: Detected article}
    \usepackage[varg]{txfonts} 
    \usepackage{tgtermes} 
    \else 
    \typeout{Fonts: Detected asy for beamer}
    \usepackage{tgheros}
    
    \fi
  }
\usepackage{microtype} 
\def\MT@register@subst@font{
  \MT@exp@one@n\MT@in@clist\font@name\MT@font@list
  \ifMT@inlist@\else\xdef\MT@font@list{\MT@font@list\font@name,}\fi}
\makeatother

\usepackage{graphicx} %
\usepackage[x11names,svgnames,rgb]{xcolor} %
\usepackage{xspace}
\usepackage{braket}
\usepackage{accents}
\usepackage{siunitx}
\usepackage{mathtools}
\DeclareSymbolFontAlphabet{\mathrm}{operators}

\definecolor{CiteColor}{rgb}{0.18039, 0.18824, 0.57255}
\definecolor{UrlColor} {rgb}{0.741, 0.173, 0.000}
\definecolor{DarkUrlColor} {rgb}{0.500, 0.110, 0.000}
\definecolor{LinkColor}{rgb}{0.25098, 0.47843, 0.04706}

\makeatletter %
\newcommand{\ShowFont}{%
  \typeout{The main font is \f@encoding \space \f@family \space %
    \f@series \space \f@shape \space at \f@size pt.}%
  \typeout{The math font sizes are \tf@size pt (main), \sf@size pt %
    (script), and \ssf@size pt (scriptscript).}%
  \typeout{The linewidth is \the\linewidth}} %
\makeatother %

\usepackage{xargs}

\usepackage{graphicx}
\usepackage{dcolumn}
\usepackage{bm}
\usepackage{tabularx}
\usepackage{enumerate}
\usepackage{multirow}
\usepackage{capt-of}
\usepackage{color}
\usepackage{url}
\usepackage[utf8]{inputenc}
\usepackage{placeins}
\usepackage{soul}

\usepackage[nolist,nohyperlinks]{acronym}
\usepackage{xspace}
\usepackage[colorlinks]{hyperref}

\usepackage{orcidlink}

\usepackage{subfigure}

\DeclareMathAlphabet{\mathbfsf}{\encodingdefault}{\sfdefault}{bx}{sl}

\newcommand{\be}{\begin{equation}}
\newcommand{\ee}{\end{equation}}
\newcommand{\bea}{\begin{eqnarray}}
\newcommand{\eea}{\end{eqnarray}}

\definecolor{light-gray}{gray}{0.95}

\definecolor{dodgerblue}{HTML}{1E90FF}
\definecolor{viennared}{HTML}{DA0A14}
\definecolor{ctorange}{HTML}{FF6C0C}
\definecolor{granadagreen}{HTML}{078931}
\definecolor{wales}{HTML}{ff0038}
\definecolor{valenciacfred}{HTML}{ee3524}
\definecolor{barcelonafcgold}{HTML}{edbb00}
\definecolor{jam}{HTML}{A50B5E}
\definecolor{austriawien}{HTML}{441678}

\AtBeginDocument{%
  \hypersetup{
    citecolor=dodgerblue,
    linkcolor=dodgerblue,   
    urlcolor=dodgerblue}}

\newcommand{\UIB}{Departament de F\'isica, Universitat de les Illes Balears, IAC3 -- IEEC, Carretera de Valldemossa km 7.5, E-07122 Palma, Spain}

\newcommand{\ICE}
{Institut de Ci\`encies de l'Espai (ICE, CSIC), Campus UAB, Carrer de Can Magrans s/n, 08193 Cerdanyola del Vall\`es, Spain}

\allowdisplaybreaks[1]

\begin{document}

\title[ML]
{A waveform model for the missing quadrupole mode  from black hole coalescence: memory effect and ringdown of the $(\ell=2,m=0)$ spherical harmonic}


\author{Maria Rossell\'o-Sastre\,\orcidlink{0000-0002-3341-3480}}
\affiliation{\UIB}

\author{Sascha Husa\,\orcidlink{0000-0002-0445-1971}}
\affiliation{\ICE}
\affiliation{\UIB}

\author{Sayantani Bera\,\orcidlink{0000-0003-0907-6098}}
\affiliation{\UIB}

\date{\today}

\begin{abstract}
In this paper we describe a model for the 
$(\ell=2, m=0)$ spherical harmonic mode of the gravitational wave signal emitted by the coalescence of binary black holes, in particular, spin-aligned systems.
This mode can be viewed as consisting of two components, gravitational wave memory and quasinormal ringdown, which are both included in our model. 
Depending on the parameters of the binary and the sensitivity curve of the detector, but in particular for high masses, the ringdown part can contribute significantly to the signal-to-noise ratio.
The model is constructed using the methods of the phenomenological waveforms program, and is calibrated to public numerical relativity data from the Simulating eXtreme Spacetimes (SXS) waveforms catalog, with the analytical results derived from the Bondi-Metzner-Sachs (BMS) balance laws. 
The code has been implemented as an extension to the computationally efficient {\tt IMRPhenomTHM} model,
it can therefore be used for computationally expensive applications such as Bayesian parameter estimation.
The region of validity of our model in the parameter space is given by: $q\leq10$ and $\chi_{1},\chi_{2}\in[-1,1]$, and no restrictions apply in terms of the length of the waveforms.
\end{abstract}


\maketitle


\section{Introduction}
\label{sec:Introduction}

The development of waveform models for analyzing the transient gravitational wave signals emitted by the coalescence of compact binaries has been driven by observational priorities, proceeding from dominant effects toward subdominant ones. 
As the sensitivity of gravitational wave detectors continues to improve, the demand for more precise waveform models grows, requiring models that account for the entirety of gravitational wave behavior, which will enable a more accurate determination of the source properties.
Initial models \cite{Ajith:2009bn,Husa:2015iqa,Khan:2015jqa,Taracchini:2012ig,Taracchini:2013rva} only described the $(\ell=2, m=\vert 2\vert)$ mode from nonprecessing quasicircular binaries. Then
the $(\ell=2, m=\vert 2\vert)$ mode in a coprecessing frame for precessing systems (see e.g.~\cite{Hannam:2013oca,Estelles:2020osj,Ramos-Buades:2023ehm, Varma:2019csw, Gamba:2021ydi}), and eventually further harmonics were added \cite{Babak:2016tgq,Ossokine_2020,Pratten:2020ceb,Garcia-Quiros:2020qpx,Estelles:2021gvs}. Another theme of increasing interest is the extension to include orbital eccentricity \cite{Ramos-Buades:2019uvh,Liu:2021pkr,Ramos-Buades:2021adz, Nagar:2021gss,Knee:2022hth}.
In this paper we describe a model for the $(\ell=2, m=0)$ spherical harmonic mode, which is the last of the quadrupole modes $(\ell=2)$ to be included in waveform models, aside from contributions that arise in precessing waveforms as coordinate frame effects \cite{Schmidt:2010it}.

We can view the observable signal of the $(2,0)$ mode as consisting of two components, a secular nonoscillatory growth corresponding to gravitational wave memory, and an oscillatory ringdown, both of which we take into account in our work. Gravitational wave memory arises in general relativity as a nonlinear, nonoscillatory effect that manifests itself as a permanent change in the space-time metric that persists even once the gravitational wave transient has passed. The eventual observation of this effect is of fundamental interest for testing the theory of general relativity. The goal of this paper is to provide a computationally efficient model that can be used in data analysis applications to account for the complete information in the $(2,0)$ mode, which we anticipate will improve parameter estimation for the upcoming 3G ground based detectors \cite{ET,CE} and the LISA space mission \cite{LISA}, and will aid the measurement of the memory effect \cite{Goncharov:2023woe,Grant:2022bla}. Analyzing collections of events, gravitational wave memory and more  generally the $(2,0)$ mode may also be detected before the 3G/LISA era, as discussed in \cite{Lasky:2016knh,Hubner:2019sly,Boersma:2020gxx}, in future observation runs of the ground based detector network of advanced LIGO \cite{LIGOScientific:2014pky}, advanced Virgo \cite{VIRGO:2014yos}, and KAGRA \cite{2021PTEP.2021eA101A}, which will rapidly extend the set of about 90 events detected to date \cite{LIGOScientific:2018mvr,LIGOScientific:2020ibl,arxiv.2111.03606}.


Our model is constructed using the methods of the phenomenological waveforms program \cite{Hannam:2013oca, Khan:2018fmp, PhysRevD.101.024056, Pratten:2020ceb, Husa:2015iqa, Khan:2015jqa, London:2017bcn, Pratten:2020fqn, Garcia-Quiros:2020qpx, Estelles:2020twz, Estelles:2020osj, Estelles:2021gvs}, more specifically piecewise closed form expressions are used to describe
the signal with a minimal number of ``coefficients''.  The values of these coefficients are fitted across the parameter space to public numerical relativity (NR) data from the SXS waveforms catalog, and analytical results derived from the Bondi-Metzner-Sachs (BMS) balance laws.
A code that evaluates the model has been implemented as an extension to the computationally efficient {\tt IMRPhenomTHM} model \cite{Estelles:2020twz} in the LALSuite \cite{lalsuite} framework, and is foreseen to become publicly available in the future.

The model we develop at this time only accounts for binary black hole (BBH) mergers when the misalingment of the black hole spins with the orbital angular momentum of the system and the orbital eccentricity can be neglected. The effect of lifting these restrictions will be studied in future work. The model is constructed in the time domain, which simplifies the description of the memory effect, as will be discussed below.

A first inspiral-merger-ringdown model for the $(2,0)$ harmonic was already developed in 2016  \cite{Cao_2016}, however it was restricted to equal mass systems with spin components satisfying $\chi_S\geq0$. More recently, a NRSurrogate model including the memory effect was presented, named as {\tt NRHybSur3dq8\_CCE}, that uses the Cauchy-characteristic evolution \cite{Yoo:2023spi}. This hybrid model is valid for $q\leq8$ and $\chi_1,\chi_2\in[-0.8,0.8]$. The model that we present in this work covers a larger region of the parameter space extending its validity up to $q=10$ and $|\chi_{1z}|,|\chi_{2z}|=1$. Later in this work we will compare the accuracy of the model presented here with the two models previously mentioned.



The paper is organized as follows. In Sec. \ref{sec:memory_summary}, we briefly summarize the relevant basic concepts of gravitational wave memory, taking calculations derived in previous works presented in the literature. Based on that, in Sec. \ref{sec:theo_frame}, we derive an analytical expression with which one can obtain the memory effect for any waveform. In Sec. \ref{sec:model_construction}, we present the strategy we have followed to construct the model of the $(2,0)$ spherical harmonic. We analyze the accuracy of the model we have constructed by computing matches and residuals in Sec. \ref{sec:accuracy}. In Sec. \ref{sec:SNR} we perform some calculations of the signal-to-noise ratio of the waveforms including memory in order to check its detectability. Sec. \ref{sec:conclusions} summarizes  the main conclusions from our work and the prospects of future work to be done. Finally, in Appendices \ref{appendix:BMS_HM} and \ref{appendix:compmodel}, we present some additional results for completeness.

In this work geometric units will be used, i.e. $c=G=1$. In order to get the redshift associated to a luminosity distance we use the Planck 2015 cosmology \cite{Planck2015}.

\section{A brief overview of gravitational wave memory}
\label{sec:memory_summary}

Gravitational memory effects can be classified into different types. First, one can distinguish linear and nonlinear memory. The existence of linear memory has been known since the 1970s \cite{Polnarev}. This kind of memory arises from systems with unbound components, and already arises at linear order in perturbation theory. Nonlinear memory was only discovered in the 1990s: Blanchet and Damour \cite{Blanchet:1992br} and Christodoulou \cite{PhysRevLett.67.1486} realized independently that outgoing gravitational wave signals can also source a memory effect, which does contribute at linear order in perturbation theory.

Recently, a new classification was proposed, based on the BMS balance laws. The symmetry group of Minkowski spacetime is the Poincaré group, which is broken
in the presence of gravitational radiation. For the computation of the memory effect we consider curved asymptotically flat spacetimes. Their symmetry group is the BMS group, an infinite-dimensional generalization of the Poincaré group. It is known that the memory effect corresponds exactly to the difference between the change in the BMS charges and the BMS fluxes \cite{Mitman:2020pbt}. The BMS group corresponds to the Poincaré group plus an infinite number of transformations referred to as supertranslations. The extended BMS group consists of, besides supertranslations, also superrotations and superboosts; each of these three transformations corresponds to a different memory effect. The close relation between the memory effect and asymptotic symmetries, as well as the associated  soft hair theorems, has generated significant theoretical interest in this area \cite{He:2014laa,Haco:2018ske,Hawking:2016msc,Strominger:2014pwa}. The importance of BMS frame-fixing in the context of fitting QNMs to waveforms that include higher modes and memory has been discussed in \cite{MaganaZertuche:2021syq}. For a detailed review on gravitational wave memory and the BMS transformations, see \cite{Mitman:2024uss}.

Ordered by the strength of the contribution for compact binary gravitational wave signals, we can distinguish three types of memory by the underlying symmetry group:
\begin{enumerate}[(i)]
    \item The normal or displacement memory is related to supertranslations, and appears directly in the strain. It results in the change in the relative position between two freely falling, initially comoving observers.
    \item The spin memory is related to superrotations, and appears in the retarded time integral of the strain. It is related to the change in the relative time delay between two freely falling observers on initially counter-orbiting trajectories.
    \item The center-of-mass memory which is related to superboosts and appears in the retarded time integral of the strain as well. It is related to the change in the relative time delay of two freely falling observers in antiparallel trajectories.
\end{enumerate}

These three types of memory generally include both contributions from linear and nonlinear components, which are also referred to as \textit{ordinary} and \textit{null} memory, respectively. This new nomenclature is used in order to avoid potential confusion about which terms are included in each memory contribution as, depending on the perturbation theory that one considers, a particular effect could appear linearly or nonlinearly.

\begin{figure} [htp]
    \includegraphics[width=0.9\columnwidth]{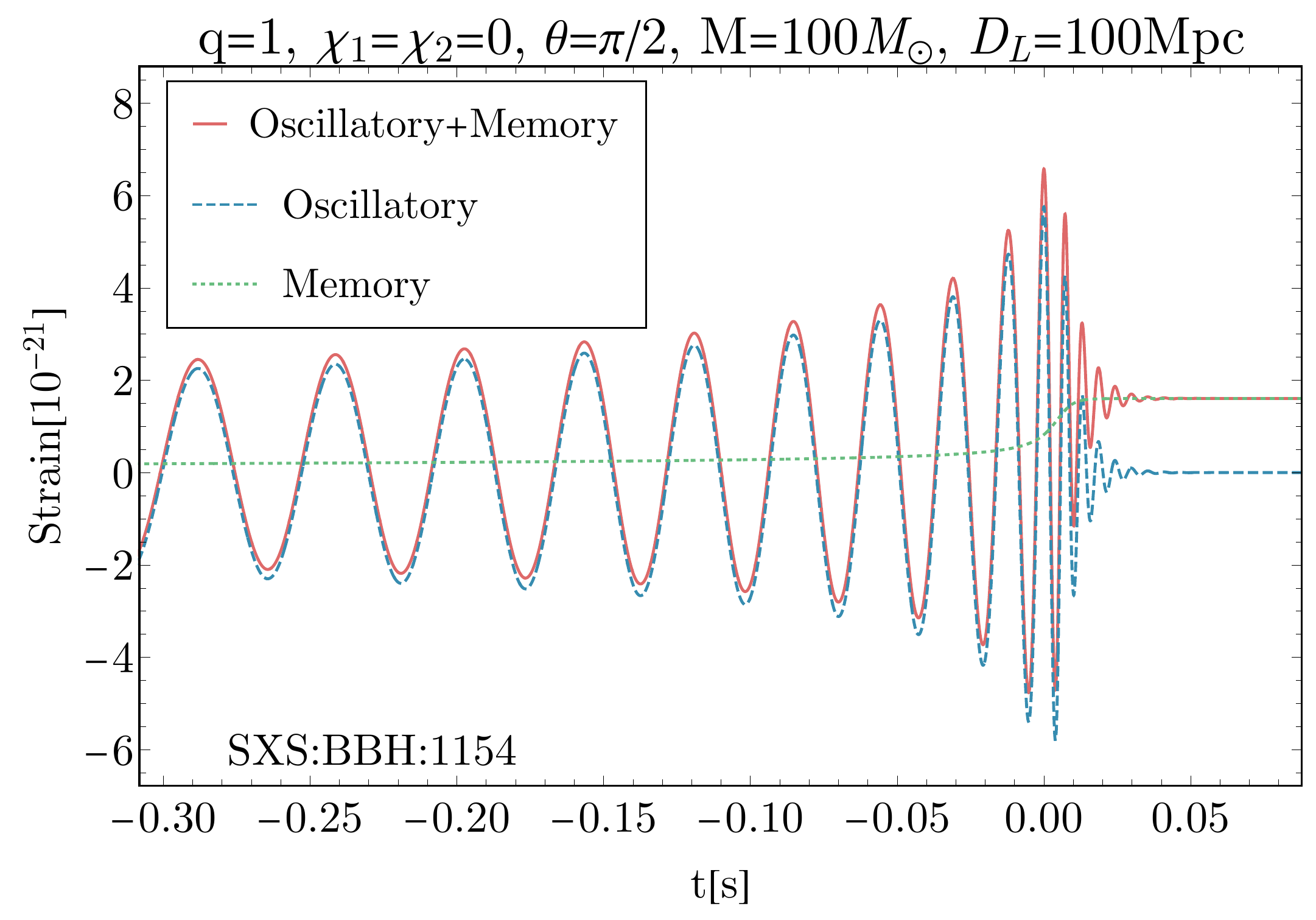}
    \caption{Comparison of the waveform for a BBH system with the parameters stated in the label of the plot, including and not including the memory effect.
    \label{fig:mem_nomem}
    }
\end{figure}

For BBH mergers the most important contribution is the displacement memory and this is the one that we are modeling in this work. Hence, from hereon when we mention the memory we refer just to the displacement memory. This effect appears as a contribution to the amplitude of the wave that grows slowly during the inspiral and suddenly rapidly near the merger. It leaves a permanent off-set that remains once the wave has passed. Therefore, the difference in the perturbation of the metric at late and early times is different from zero for at least one of the two polarizations \cite{Favata:2010zu},
\begin{equation}
    \Delta h_{+,\times}^{\text{(mem)}}= \lim\limits_{t \to +\infty}h_{+, \times}(t)-\lim\limits_{t \to -\infty}h_{+, \times}(t)\neq0.
\end{equation}

In Fig. \ref{fig:mem_nomem} we present a signal for an equal mass, nonspinning system (SXS:BBH:1154) with (red, solid) and without memory (blue, dashed).

Displacement memory is maximal for equal mass binaries with large spins aligned with the orbital angular momentum, as they are the ones with the highest energy flux \cite{Pollney:2010hs}. The $(2,0)$ harmonic gives a maximal contribution for edge-on systems and vanishes for face-on, unlike the dominant contribution from the $(2, \pm2)$ mode, which is maximal when viewed  face-on or face-off \cite{Talbot:2018sgr,Mitman:2022qdl}.
This feature can help to break the distance-inclination degeneracy, when its signal-to-noise ratio is sufficiently high, and it could also help to constrain further BBH parameters. See \cite{xu2024enhancing} for a recent work on this topic using the upgraded LIGO-A$^{\#}$ sensitivity and \cite{Gasparotto:2023fcg} for a case study for the LISA mission. Both works consider only the memory contribution, but not the oscillatory ringdown signal of the $(2,0)$ mode that we consider here. Moreover, memory would also be useful to distinguish BBH from NSBH mergers, as studied in \cite{Lopez:2023aja,Tiwari:2021gfl}.

\section{Theoretical framework}
\label{sec:theo_frame}

\subsection{Spherical harmonic decomposition}
\label{subsec:spherical}

The discussion of the memory effect is greatly simplified by employing the standard decomposition of the gravitational wave strain in a spherical harmonic basis of spin weight -2:
\begin{equation}
    h(t,r,\theta,\phi)=h_+-ih_{\times}=\frac{1}{r}\sum_{\ell=2}^{\infty}\sum_{m=-\ell}^{m=\ell}h_{\ell,m}(t)\;^{-2}Y_{\ell,m}(\theta,\phi),
\end{equation}
where
\begin{equation}
    ^sY_{\ell,m}(\theta,\phi)=(-1)^s\sqrt{\frac{2\ell+1}{4\pi}}\;^{-s}d_{\ell,m}(\theta)e^{im\phi},
\end{equation}
and the angular coordinates are related to Cartesian coordinates in the standard way.
For nonprecessing binaries, the orbital plane is conserved, and we consider a binary orbiting in the $x-y$ plane, with the 
angular momentum parallel or antiparallel to the $z$ axis. This way $\theta$ is the angle between the angular momentum vector and the line of sight between the binary and the detector; and $\phi$ is the angle between the $x$ axis and the line of sight to the detector projected onto the plane of the binary.

For a nonprecessing binary, the modes that contain the highest contributions of the memory are the ones with $m=0$. But, as shown in \cite{Talbot:2018sgr}, including the $(3,\pm3)$ mode in the spherical harmonic decomposition of the strain, gives rise to a small contribution to the second polarization for systems with asymmetric components (nonvanishing $(2,\pm1)$ mode). However, the main contributions are contained in the  (2,0) and (4,0) modes, where the contribution in the $\ell=2$ is two orders of magnitude larger. For this reason, this is the spherical harmonic for which we construct the model. 

\subsection{Null infinity}
\label{subsec:bondi}

The  outgoing gravitational wave signal is 
only unambiguously defined at future null infinity (see e.g. \cite{Wald_textbook, conformalinf}), which idealizes observers of radiation phenomena at astronomical distances.
Computing the gravitational wave signal close to the source, and then extrapolating to null infinity has been very (maybe surprisingly) successful in numerical relativity, however this is not sufficient to compute the memory effect. Numerical codes that use Cauchy-Characteristic extraction, where the Einstein equations are evolved along outgoing null surfaces far from the source have however been able to compute memory signals numerically \cite{Pollney:2010hs, SXSCCEarticle}.  See however \cite{Giannakopoulos:2023zzm} for subtleties with the well-posedness of such formulations.

The Bondi framework provides the mathematical tools to compute physical quantities at future null infinity, which can be thought to be the limiting surface at $r=\infty$ and it is obtained when moving away from sources along null surfaces of constant retarded time. This is the region of space-time where observable quantities as the strain of the gravitational waves are defined in a nonambiguous way. The procedure to get such quantities consists of foliating the space-time by outgoing null hypersurfaces of constant retarded time which are again foliated by space-like 2-spheres of constant $r$. Then, one can obtain the Weyl-Newman-Penrose scalars. To check the detailed deduction, see \cite{Ashtekar:2019viz}. All the five scalars have a physical interpretation but of special interest is the scalar $\Psi_4$, as it is the one that describes the transverse radiation and characterizes the outgoing gravitational radiation. Hence, the two degrees of freedom of gravitational waves (or polarizations) can be obtained from $\Psi_4=-\Ddot{h}_++i\Ddot{h}_{\times}$.

\subsection{Eth formalism}
\label{subsec:eth_formalism}

The eth ($\eth$) formalism provides an efficient way of treating tensor fields and their covariant derivatives on the sphere \cite{ethformalism}.
In order to compute the quantities we are interested in Sec. \ref{subsec:BMS_calculation}, we need to know how the spin-weighted derivative operators act on spin-weighted spherical harmonics,
\begin{equation}
\eth^sY_{\ell,m}=+\sqrt{(\ell-s)(\ell+s+1)}\; ^{s+1}Y_{\ell,m},
\end{equation}
\begin{equation}
\label{ethbar}
\bar{\eth}^sY_{\ell,m}=-\sqrt{(\ell-s)(\ell-s+1)}\; ^{s-1}Y_{\ell,m}.
\end{equation}
One can also construct the spherical Laplacian $D^2$ which, when acting on $s=0$ spin-weight spherical harmonics, can be defined as:
\begin{equation}
\label{mathd}
D^2Y_{\ell,m}=\bar{\eth}\eth Y_{\ell,m} =\eth\bar{\eth} Y_{\ell,m}=-\ell(\ell+1)Y_{\ell,m}.
\end{equation}
For the calculations of the BMS balance laws discussed in the next subsection,  it will be useful to define the operator $\mathfrak{D}$ as
\begin{equation}
\label{eq:defineD}
\mathfrak{D}Y_{\ell,m}=\left[\frac{1}{8}D^2(D^2+2)\right]Y_{\ell,m}=\frac{1}{8}(\ell+2)(\ell+1)\ell(\ell-1)Y_{\ell,m}.
\end{equation}

\subsection{BMS balance laws}
\label{subsec:BMS}
In this subsection, we examine the relation between the BMS balance laws and memory, as was discussed in \cite{Mitman:2020bjf, Mitman:2024uss}. Displacement memory can be obtained from the supermomentum balance law
\begin{equation}
    \eth^2h=\int_{-\infty}^u\left|\dot{h}\right|^2du-4\left(\Psi_2+\frac{1}{4}\dot{h}\bar{h}\right)-4M_{\text{ADM}}.
\end{equation}
where $M_{\text{ADM}}$ is the corresponding  Arnowitt-Deser-Misner (ADM) mass, $u$ is the retarted Bondi time and $\Psi_2$ is the Weyl scalar. \\
This equation can be solved for by writing it as \cite{Mitman:2020bjf}
\begin{equation}
\label{hbondi}
    h=\frac{1}{2}\bar{\eth}^2\mathfrak{D}^{-1}\left[\frac{1}{4}\int_{-\infty}^u\left|\dot{h}\right|^2du-\left(\Psi_2+\frac{1}{4}\dot{h}\bar{h}\right)\right],
\end{equation}
where the operator $\mathfrak{D}$ has been defined in Eq.~(\ref{eq:defineD}), and $\mathfrak{D}^{-1}$ is its inverse. We are working with an integro-differential equation, needing to invert a differential operator, which may seem nontrivial. But, as we are decomposing the strain in a spherical harmonic basis, as explained in Sec. \ref{subsec:spherical}, it is sufficient to invert the analytical expression of $\mathfrak{D}$, so that, acting on $Y_{\ell,m}$, the operator is defined as
\begin{equation}
    \mathfrak{D}^{-1}Y_{\ell,m}=\frac{8}{(\ell+2)(\ell+1)\ell(\ell-1)}Y_{\ell,m}.
\end{equation}
Going back to Eq. (\ref{hbondi}), it is clear that one does not have access to the past infinity, but rather to a finite starting time, $u_1$. Also, since, there might be some discrepancies regarding the post-Newtonian Bondi frame, the true strain ($h$) and the strain computed from the BMS balance laws ($J$) can differ by an unknown constant ($\alpha$), which depends on the angular coordinates and is related to supertranslations. Taking all these into consideration, one can write
\begin{equation}
\label{Jalpha}
    J\equiv\frac{1}{2}\bar{\eth}^2\mathfrak{D}^{-1}\left[\frac{1}{4}\int_{u_1}^u\left|\dot{h}\right|^2du-\left(\Psi_2+\frac{1}{4}\dot{h}\bar{h}\right)\right]+\alpha(\theta,\phi).
\end{equation}
This quantity $J$ can be split into four different terms that are directly related to the memory contributions,
\begin{equation}
\label{Jsum} J=J_m+J_{\varepsilon}+J_{\widehat{N}}+J_{\mathcal{J}},
\end{equation}
in terms of common physical quantities their interpretations are: the Bondi mass aspect, the energy flux, the angular momentum aspect and the angular momentum flux, respectively. As one can check in \cite{Mitman:2020bjf}, the most prominent contribution to memory is provided by the energy flux term, $J_{\varepsilon}$. Therefore we will calculate the memory by taking into account only this term, which has the following expression
\begin{equation}
\label{je}
    J_{\varepsilon}=\frac{1}{2}\bar{\eth}^2\mathfrak{D}^{-1}\left[\frac{1}{4}\int_{u_1}^u\left|\dot{h}\right|^2 du\right]+\alpha.
\end{equation}

\subsection{Using the BMS balance laws to compute memory}
\label{subsec:BMS_calculation}
In this section we reproduce some calculations that have already been presented in previous works. We write down the results in order to provide the full theoretical framework in which are working and to facilitate the reader the understanding of our procedure for computing the memory contribution.

Starting from Eq. (\ref{je}), we aim to obtain an expression that gives us the relation between the memory and the modes of the gravitational wave strain.  In this way,  one is able to calculate the memory related to any waveform using the $(\ell, m)$ modes. The code implemented in \cite{sxs_mem} gives the memory correction for the simulations in the SXS Catalog, i.e. in the SXS format. In order to be able to compute the memory contribution for any waveform, we present analytical expressions so one does not need to rely on this specific format to carry out the calculations. Previous works in the literature provide this result but only including the dominant quadrupolar $(2,\pm2)$ modes, see e.g. \cite{Favata_2009}. In the following, we present the mathematical deduction from the BMS balance laws for this case. And, in addition, we carry out the same procedure including higher modes. In this way, we present a new and simple way to get the memory contribution more accurately through analytical expressions.

As we have introduced, as a first approach,  we will only consider the dominant quadrupolar modes $(2,\pm2)$ as they are roughly an order of magnitude larger than the rest of the modes.\\ 
Neglecting higher modes,  we can approximate the strain by
\begin{equation}
    h(t,\theta,\phi)\approx h_{2,2}(t)\;^{-2}Y_{2,2}(\theta,\phi)+h_{2,-2}\;^{-2}Y_{2,-2}(\theta,\phi).
\end{equation}
We need to calculate $\left|\dot{h}\right|^2$,  where the over-dot indicates the time derivative,
\begin{equation}
    \dot{h}(t,\theta,\phi)\approx \dot{h}_{2,2}(t)\;^{-2}Y_{2,2}(\theta,\phi)+\dot{h}_{2,-2}(t)\;^{-2}Y_{2,-2}(\theta,\phi).
\end{equation}
We then rewrite the mode strain in terms of its real and imaginary parts explicitly and we use the relation $h_{\ell, m}=(-1)^{\ell}h^*_{\ell, -m}$, which is also valid for the time derivative, to write:
\begin{equation}
\begin{split}
    &\dot{h}_{2,2}(t)= \text{Re}\left[\dot{h}_{2,2}(t)\right]+i\;\text{Im}\left[\dot{h}_{2,2}(t)\right],\\
    &\dot{h}_{2,-2}(t)= \text{Re}\left[\dot{h}_{2,2}(t)\right]-i\;\text{Im}\left[\dot{h}_{2,2}(t)\right].
\end{split}
\end{equation}
To simplify the notation, we define $\dot{h}_{2,2}^{\text{Re}}\equiv\text{Re}\left[\dot{h}_{2,2}(t)\right]$ and $\dot{h}_{2,2}^{\text{Im}}\equiv\text{Im}\left[\dot{h}_{2,2}(t)\right]$.  Using the spherical harmonic
\begin{equation}
\label{Y22}
    ^{-2}Y_{2,\pm2}(\theta,\phi)=\sqrt{\frac{5}{64\pi}}(1\pm\cos\theta)^{2}e^{\pm2i\phi},
\end{equation}
and computing the squared absolute value of $\dot{h}$, we get
\begin{equation}
\begin{split}
    |\dot{h}|^2=\dot{h}\dot{h}^*\approx&\frac{5}{256\pi} \left\{8 \sin ^4\theta \left[\left(\dot{h}_{2,2}^{\text{Re}}\right)^2-\left(\dot{h}_{2,2}^{\text{Im}}\right)^2\right] \cos (4 \phi )\right.\\
    &+\left[28 \cos (2 \theta )+\cos (4 \theta )+35\right] \left[\left(\dot{h}_{2,2}^{\text{Re}}\right)^2+\left(\dot{h}_{2,2}^{\text{Im}}\right)^2\right]\\
    &\left.-16\;\dot{h}_{2,2}^{\text{Re}}\;\dot{h}_{2,2}^{\text{Im}} \sin ^4\theta \sin (4 \phi )\right\}.
\end{split}
\end{equation}

To apply the operators $\mathfrak{D}^{-1}$ and $\bar{\eth}^2$,  we need to project this quantity onto the $s=0$ spherical harmonics 
\begin{equation}
    \int_{\phi=0}^{2\pi}\int_{\theta=0}^{\pi}\; ^0Y_{2,0}(\theta)\; \left|\dot{h}\right|^2 \sin\theta\; d\theta d\phi=\frac{2}{7}\sqrt{\frac{5}{\pi}}\left|\dot{h}_{2,2}\right|^2,
\end{equation}
\begin{equation}
    \int_{\phi=0}^{2\pi}\int_{\theta=0}^{\pi}\; ^0Y_{4,0}(\theta)\; \left|\dot{h}\right|^2 \sin\theta\; d\theta d\phi=\frac{\left|\dot{h}_{2,2}\right|^2}{42\pi}.
\end{equation}
Hence,  in terms of these two spherical harmonics, we have
\begin{equation}
    \left|\dot{h}\right|^2\approx \left|\dot{h}_{2,2}\right|^2\left[\frac{2}{7}\sqrt{\frac{5}{\pi}}\;^0Y_{2,0}(\theta)+\frac{1}{42\pi}\;^0Y_{4,0}(\theta)\right].
\end{equation}
Now,  we have to apply the two operators to this last result. The action of $\bar{\eth}$ on the spherical harmonics is given in Eq. (\ref{ethbar}).  It is important to notice that $\bar{\eth}^2$ decreases the value of $s$ by two.  The expression for $\mathfrak{D}^{-1}$ can be deduced from (\ref{mathd}).  Then,  replacing the expression of the $s=-2$ spherical harmonics,
\begin{equation}
\label{Y20}
    ^{-2}Y_{2,0}(\theta)=\frac{1}{4}\sqrt{\frac{15}{2\pi}}\sin^{2}\theta ,
\end{equation}
\begin{equation}
\label{Y40}
    ^{-2}Y_{4,0}(\theta)=\frac{3}{16}\sqrt{\frac{5}{2\pi}}\left[5+7\cos(2\theta)\right]\sin^{2}\theta,
\end{equation}
we get
\begin{equation}
\begin{split}
\label{operators22}
\frac{1}{8}\bar{\eth}^2\mathfrak{D}^{-1} \left|\dot{h}\right|^2&\approx \left|\dot{h}_{2,2}\right|^2\left[\frac{1}{7} \sqrt{\frac{5}{6 \pi }}\; ^{-2}Y_{2,0}(\theta)+\frac{1}{252 \sqrt{10 \pi }}  \;^{-2}Y_{4,0}(\theta)\right]\\
    &=\left|\dot{h}_{2,2}\right|^2\frac{(17+\cos^2\theta)\sin^2\theta}{192\pi}.
\end{split}
\end{equation}
By recovering the time integral, we get the full expression to compute the memory when we only consider the dominant modes,
\begin{equation}
J_{\varepsilon}\approx\frac{(17+\cos^2\theta)\sin^2\theta}{192\pi}\int_{u_1}^u\left|\dot{h}_{2,2}\right|^2\;du,
\end{equation}
which is the same analytic expression as in \cite{Favata_2009}. We are interested in the memory component of the $(2,0)$ mode, hence, from Eq. (\ref{operators22}), it is easy to see that the projection onto this spherical harmonic gives
\begin{equation}
\label{h20mem22}
h^{\text{(mem)}}_{2,0}=\frac{1}{7}\sqrt{\frac{5}{6\pi}}\int_{u_1}^u\left|\dot{h}_{2,2}\right|^2\;du.
\end{equation}

\begin{center}
\begin{figure}[htp]
\includegraphics[width=0.9\columnwidth]{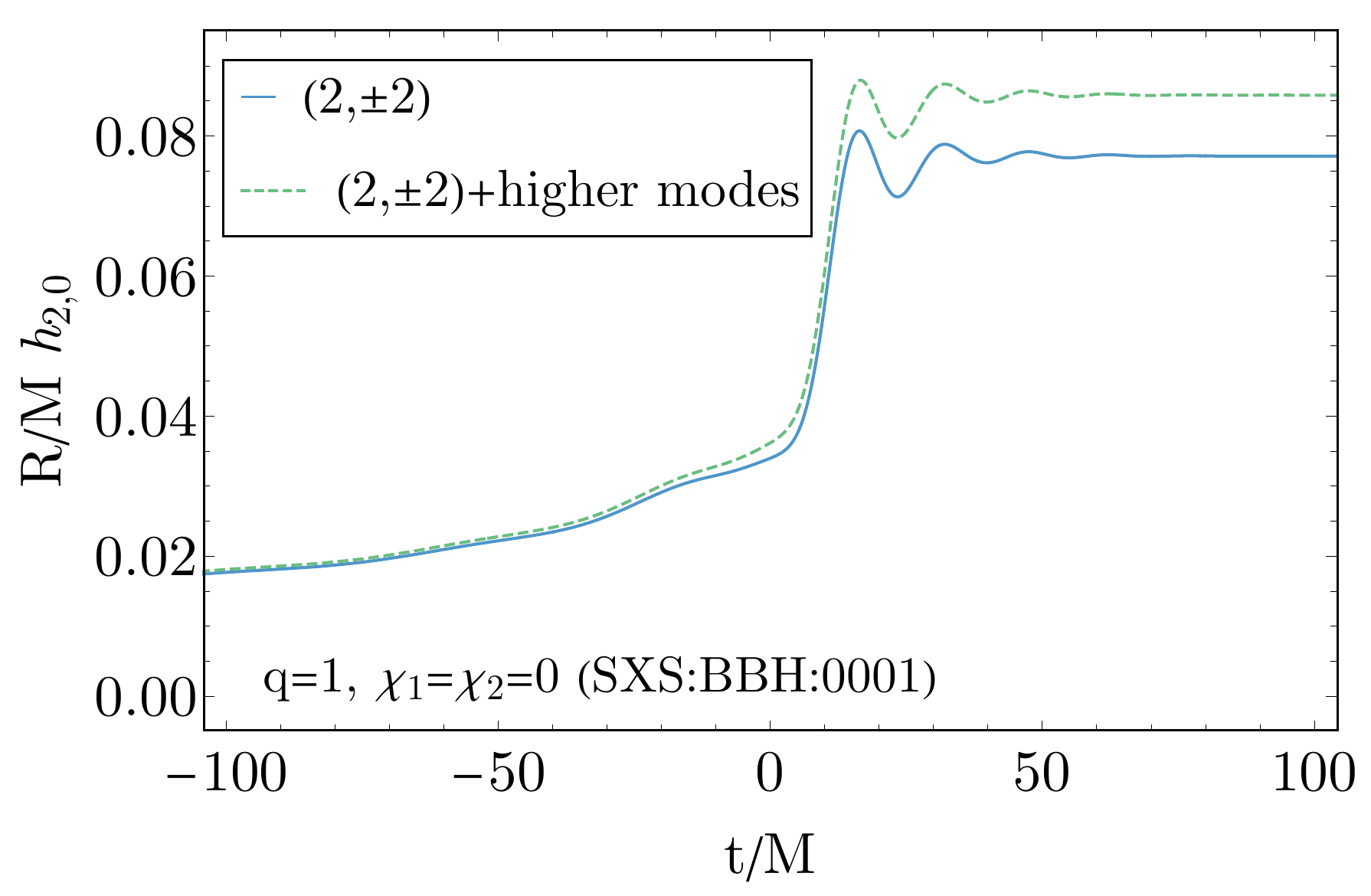}
    \caption{$(\ell=2,m=0)$ spherical harmonic with the memory contribution computed using Eq. (\ref{h20mem22}) (blue curve) and using Eq. (\ref{h20memhm}) (green, dashed curve) for an equal mass, nonspinning system. No shift is performed on these waveforms, therefore $h_{2,0}(f=0)\neq0$.
    \label{fig:Flmexample}
    }
\end{figure}
\end{center}

We carry out the analogous procedure as above, but now we also include the modes $(\ell,  m)=(2,\pm1),(3,\pm2),(3,\pm3),(4,\pm4)$ in addition to the dominant modes. Details of these calculations are provided in Appendix \ref{appendix:BMS_HM}. When computing the absolute value squared of $\dot{h}$,  one obtains an expression in terms of the modes,  which depends explicitly on the angular coordinates $\theta$ and $\phi$.  The result of $\left|\dot{h}\right|^2$ is then projected onto the basis of $s=0$ spherical harmonics. The results of the projections can be found in the Eqs. (\ref{proj20}-\ref{proj80}). Finally, one applies the corresponding operators and gets to the final result, which can be expressed in terms of the functions $F_{\ell,\pm m}(\theta)$ as follows:
\begin{equation}
\begin{split}
\label{integrand}
J_{\varepsilon}\approx&\int_{u_1}^u\left[F_{2,\pm 2}(\theta)\left|\dot{h}_{2,2}\right|^2+F_{2,\pm 1}(\theta)\left|\dot{h}_{2,1}\right|^2+F_{3,\pm 2}(\theta)\left|\dot{h}_{3,2}\right|^2\right.\\
    &+F_{2,\pm2;3,\pm 2}(\theta)\left(\dot{h}_{2,2}^{\text{Re}}\dot{h}_{3,2}^{\text{Re}}+\dot{h}_{2,2}^{\text{Im}}\dot{h}_{3,2}^{\text{Im}}\right)+
F_{3,\pm 3}(\theta)\left|\dot{h}_{3,3}\right|^2\\
&\left.+F_{4,\pm 4}(\theta)\left|\dot{h}_{4,4}\right|^2\right]\; du.
\end{split}
\end{equation}
Here we notice that the modes $(2,\pm2)$ and $(3,\pm2)$ give a mixed term. The function $F_{\ell,\pm m}(\theta)$ depends only on the inclination $\theta$, and the explicit forms are presented in equations (\ref{F22}-\ref{F44}). Again, we are interested in the projection onto the $(2,0)$ spherical harmonic in order to obtain the specific contribution to this mode, which gives
\begin{equation}
\begin{split}
\label{h20memhm}
&h_{2,0}^{\text{(mem)}}=\frac{1}{7}\sqrt{\frac{5}{6\pi}}\int_{u_1}^u\left|\dot{h}_{2,2}\right|^2\;du-\frac{1}{14}\sqrt{\frac{5}{6\pi}}\int_{u_1}^u\left|\dot{h}_{2,1}\right|^2\;du\\
&+\frac{5}{2\sqrt{42\pi}}\int_{u_1}^u\left(\dot{h}_{2,2}^{\text{Re}}\dot{h}_{3,2}^{\text{Re}}+\dot{h}_{2,2}^{\text{Im}}\dot{h}_{3,2}^{\text{Im}}\right)du-\frac{2}{11}\sqrt{\frac{2}{15\pi}}\int_{u_1}^u\left|\dot{h}_{4,4}\right|^2\;du.
\end{split}
\end{equation}

At this point, we compare the waveforms of the $(2,0)$ mode that we obtain when using only the dominant modes and when including higher modes. As an example, we choose the simulation SXS:BBH:0001, presented in Fig. \ref{fig:Flmexample}. We conclude that, although the integral with the dominant modes gives an approximation of the general picture of the mode, if one does not consider the contribution of the higher modes, a fraction of the memory amplitude is lost, where the highest contribution comes from the mixed term of the $(2,\pm2)$ and $(3,\pm2)$ modes. As expected, the effect of considering higher modes does not give the same contribution across the parameter space. We find that it is more considerable when the memory amplitude is larger, therefore, for more symmetric systems (equal mass, aligned spins).

\section{Construction of the model}
\label{sec:model_construction}

\begin{figure}[htp]
    \includegraphics[width=0.8\columnwidth]{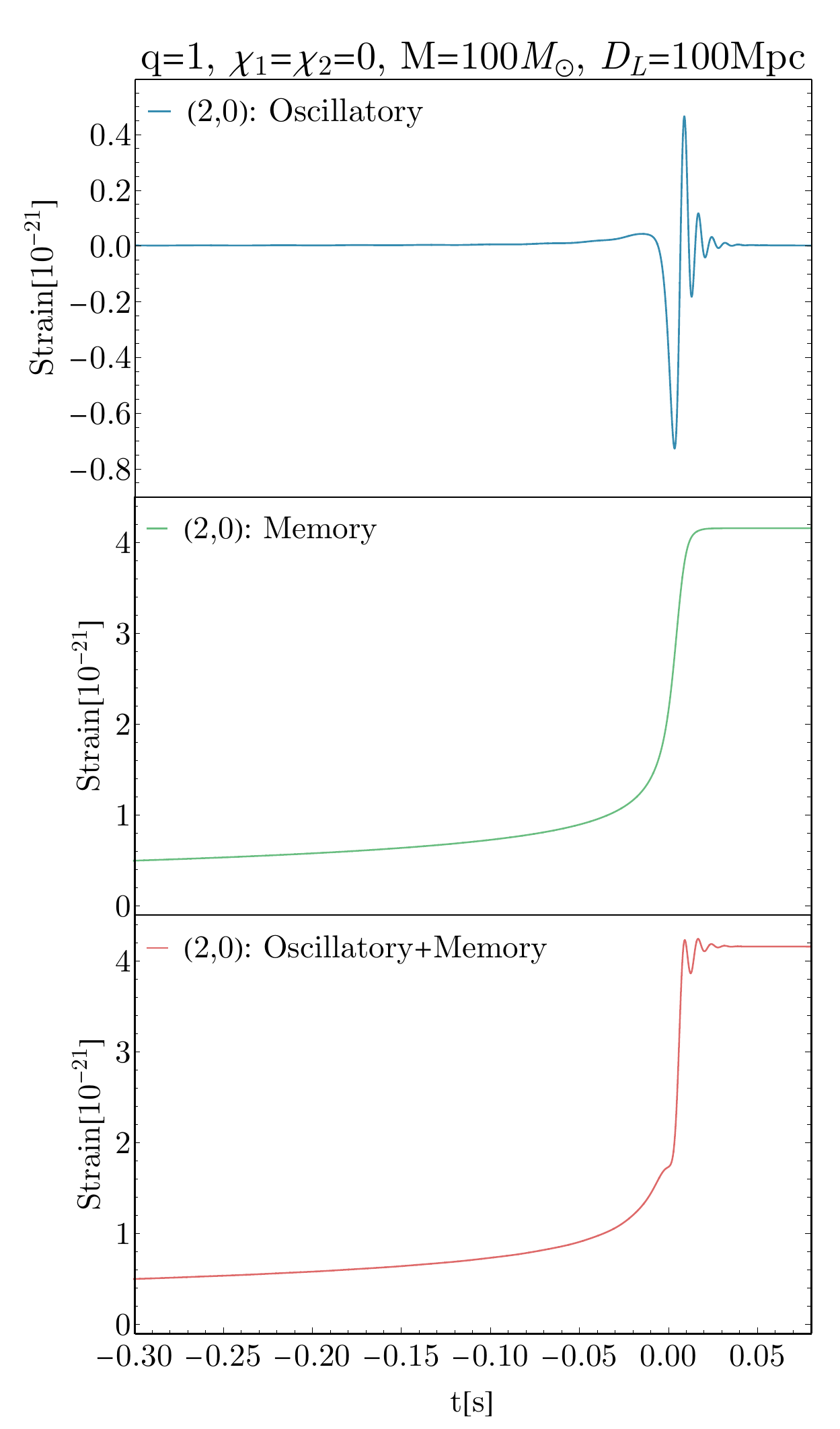}
    \caption{Waveform for an equal mass, nonspinning system (SXS:BBH:1154), using SI units. Top panel: oscillatory contribution to the $(2,0)$ spherical harmonic. Middle panel: memory contribution to the $(2,0)$ spherical harmonic. Bottom panel: full picture of the mode containing both contributions added together.
    \label{fig:20decomposed}
    }
\end{figure}

We now proceed to explain how we have constructed the model for the $(2,0)$ spherical harmonic mode, which contains  both the memory  and oscillatory ringdown contributions.
In Fig. \ref{fig:20decomposed} each individual contribution is displayed separately, together with their sum. It is clear that attempting to develop an analytical \textit{Ansatz} that accurately describes the full behavior of the curve in the bottom panel is challenging as it has subtle features to deal with. For this reason, the strategy that we follow to construct the model is to fit the oscillatory and the memory contributions separately. We separately develop different strategies for the ringdown oscillations, shown in the top panel, and for the secular growth of the memory, shown in the middle panel. Finally, we sum both contributions to obtain the full description of the $(2,0)$ spherical harmonic. Regarding the memory part, a good approach could be to directly implement the integral of the modes, but this procedure would get more complex as we need to integrate over several modes, making the model slower. For comparison, we consider the Python implementation of the memory contribution of our model and the corresponding implementation of the memory integral. For an equal-mass, nonspinning system, with $f_{\text{min}}=10$ Hz, $f_{\text{max}}=2096$ Hz and a sampling interval of 1/4096 s,  we find that the calculation of the integral, which also includes the time that takes producing the modes using the {\tt IMRPhenomXHM} model (see Sec. \ref{subsec:NR_datasets} for the explanation why {\tt IMRPhenomTHM} is not employed for this purpose), takes around 26 ms. On the other hand, the evaluation of the phenomenological \textit{Ansatz} takes around 3.4 ms. Consequently, the overall evaluation of the model would be slowed down approximately 7.5 times for this simple choice of the parameters. Moreover, when working with future detectors, the waveforms become much longer and it is important to be able to parallelize the calculations. Analytical \textit{Ansätze} are straightforward to compute in a GPU, whereas the integral of the modes would be more difficult to parallelize. For these reasons, we decided to rely on closed-form, piecewise analytical expressions which do not depend on the mode content and are fast to evaluate, making the model more computationally efficient.

\subsection{Calibration datasets}
\label{subsec:NR_datasets}

For the calibration of the oscillatory contribution and the inspiral regime of the memory contribution we rely on NR simulations that provide an accurate  $(2,0)$ mode. The convention that we consider for the mass ratio is such that $q=m_1/m_2\geq1$ and we shift the waveforms so that $t=0$ corresponds to the time of the merger determined by the peak of the amplitude of the $(2,2)$ mode. Two different sets of NR simulations are used:
\begin{itemize}
    \item SXS Catalog \cite{SXS:catalog}: 438 quasicircular, nonprecessing simulations with $q\leq10,\;\chi_1,\;\chi_2\in[-0.969,0.998]$.
    \item EMRI simulations \cite{Apte:2019txp,Lim:2019xrb} : 10 quasicircular, nonprecessing simulations with $q=1000$ and $\chi=\left\{-0.9,\; -0.7,\; -0.5,\; -0.3,\; -0.1,\; 0.1,\; 0.3,\; 0.5,\; 0.7,\; 0.9\right\}$.
\end{itemize}

\begin{figure}[htp]
     \centering
     \begin{subfigure}
         \centering
         \includegraphics[width=0.75\columnwidth]{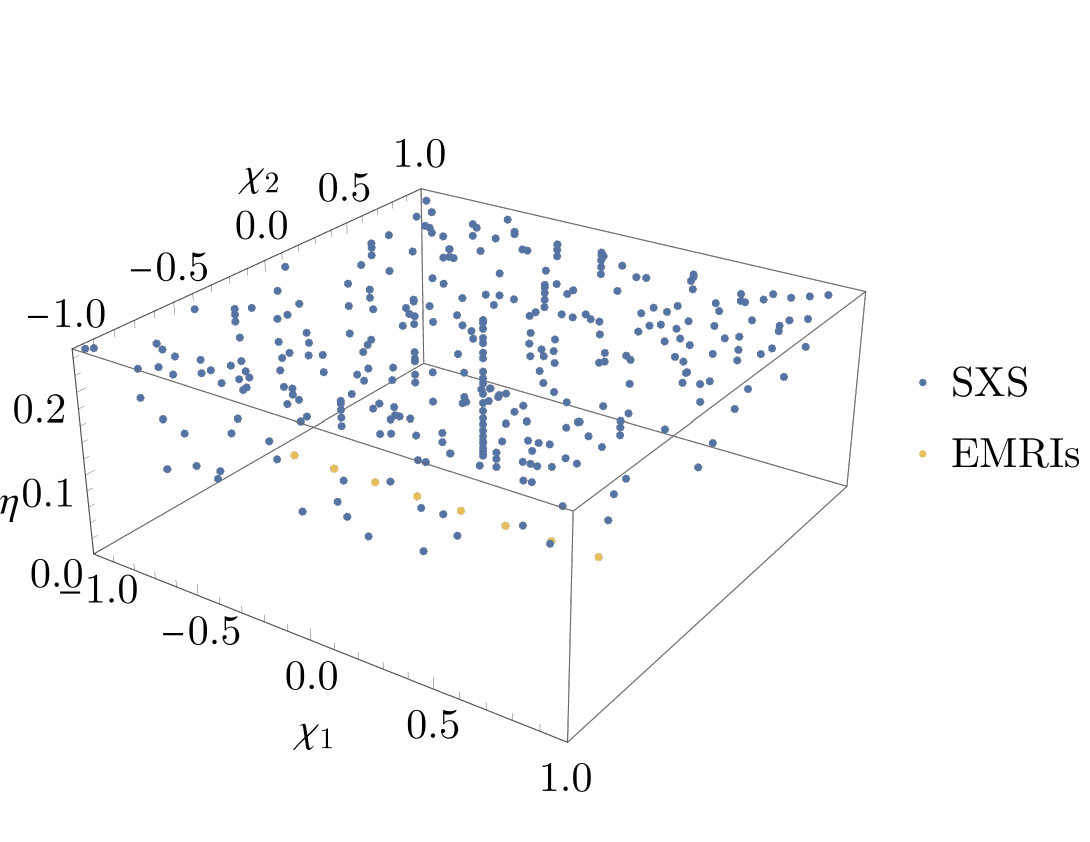}
     \end{subfigure}
     \hfill
     \begin{subfigure}
         \centering
         \includegraphics[width=0.9\columnwidth]{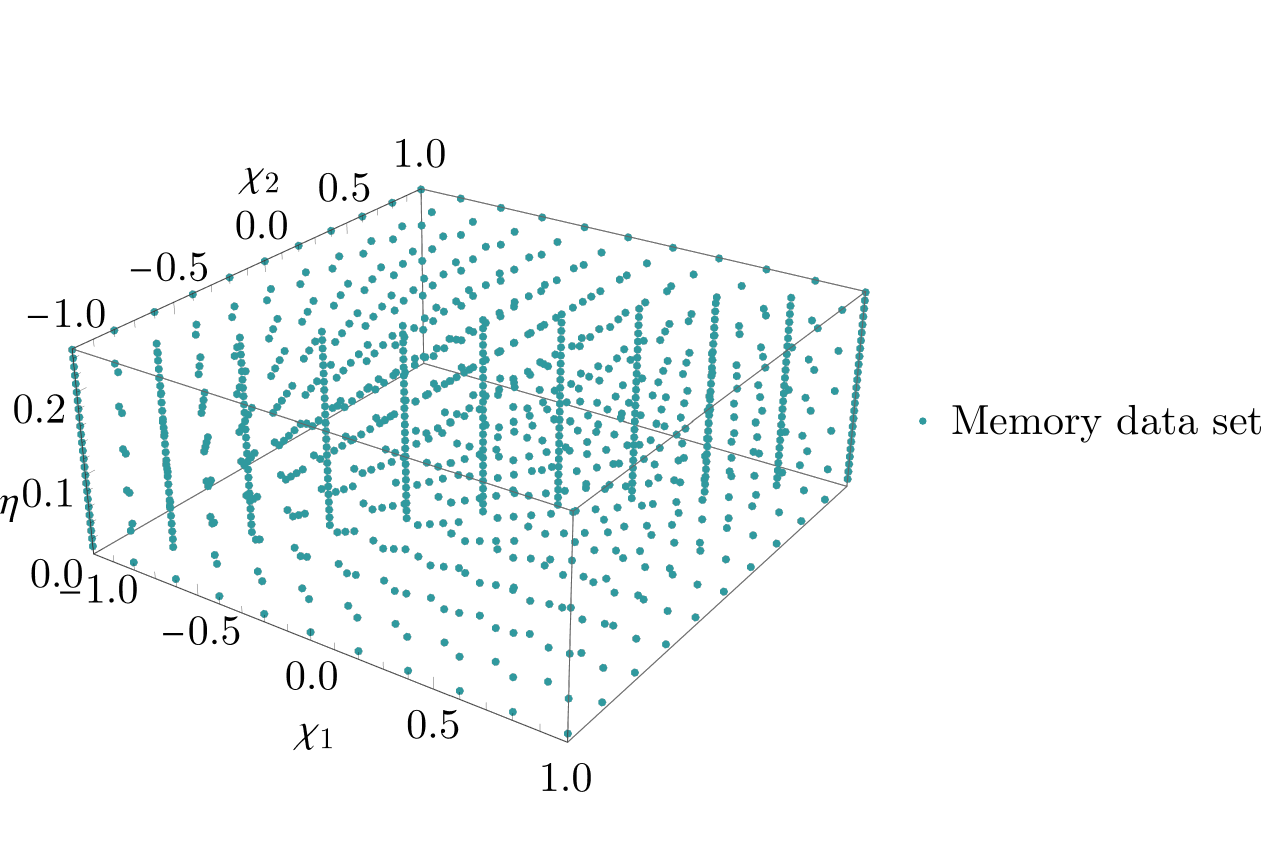}
     \end{subfigure}
        \caption{Waveforms used for the calibration of the model in the 3D parameter space $\left\{\eta,\chi_1,\chi_2\right\}$. In the top plot we present the NR simulations for the oscillatory part and the inspiral regime of the memory; and in the bottom plot, the extra dataset we created to calibrate the remaining parts of the memory contribution.}
        \label{fig:dataset}
\end{figure}

We discard a total of 7 SXS simulations, for which we believe that we cannot extract the correct physical behavior and therefore should not be included in our fits. SXS:BBH:0157, SXS:BBH:0158, SXS:BBH:0176, SXS:BBH:1124 and SXS:BBH:1146 all have $1\leq q\leq1.5$, $\chi_1,\chi_2\geq0.9$, and these high values of the individual spin components cause the amplitude of the ringdown oscillations to be so small that are masked by noise. In this region with high positive spins we find that we cannot trust any of the simulations, causing the parameter space from the SXS simulations to be reduced to $q\leq10,\;\chi_1,\;\chi_2\in[-0.969,0.900]$. SXS:BBH:1110 ($q=7,\;\chi_1=\chi_2=0$) presents some oscillations in the inspiral which we think are not physical. For SXS:BBH:1412 ($q=1.63,\;\chi_1=0.4,\;\chi_2=-0.3$), the variables of the \textit{Ansätze} fitted across the parameter space were far out from the surfaces generated by the fits of the rest of simulations. In any case, discarding these two simulations does not have any serious implication since in the dataset we have other simulations with similar parameters. In the same way, from the EMRI set of simulations, we discard the ones with $\chi=0.7$ and $\chi=0.9$ since the strain has a behavior that presents the same features as the SXS ones with high positive spins. Thus, we have 431 (SXS) + 8 (EMRIs) = 439 remaining NR simulations to calibrate our model. In the top plot of Fig. \ref{fig:dataset} we display the distribution of these simulations across the parameter space.

In the case of the memory contribution, we find that the fits across the parameter space are affected by the lack of simulations, especially for high mass ratios. We can however use the analytical expression in (\ref{h20memhm}), which can be evaluated for any mass ratio and spins. In order to get the modes that we need for this calculation, we use the phenomenological model {\tt IMRPhenomXHM} \cite{Garcia-Quiros:2020qpx}, which includes the $(\ell,|m|)=(2,2),(2,1),(3,3),(3,2),(4,4)$ modes. We choose to use this model instead of {\tt IMRPhenomTHM} because it contains the $(3,2)$ mode which we found that gives a considerable contribution to the memory part of the waveform. Although {\tt IMRPhenomXHM} is constructed in the frequency domain, we use the modes in the time domain to perform the integration. In order to provide a dense grid of data points for our fits we create a dataset a group of 935 data points, obtained by computing the memory for the {\tt IMRPhenomXHM} modes. To construct this new dataset, first, we use a uniform grid for the three parameters: $\eta=(0.01, 0.05, 0.10, 0.15, 0.20, 0.25)$ and $\chi_1\in[-1,1]$ and $\chi_2\in[-1,1]$ in steps of 0.2. Then we add a denser two-dimensional grid of equal spin data, evenly spaced: $\eta\in[0.02,0.24]$ in steps of 0.01 and shown in the bottom plot of Fig. \ref{fig:dataset}.

\subsection{Fits for the memory contribution}
\label{subsec:fits_mem}

We model the growth in amplitude of the mode sourced by the memory as represented in the left bottom panel of Fig.~\ref{fig:20decomposed}. We split this procedure in two parts, by modeling separately the inspiral regime and the late inspiral-merger-ringdown.

\subsubsection{Inspiral regime}

The inspiral regime of the memory contribution of the $(2,0)$ mode can be described by a post-Newtonian (PN) expression, which at 2.5 PN order reads \cite{Cao_2016}
\begin{equation}
\begin{split}
    \frac{R}{M}h_{2,0}^{\text{(PN,mem)}}=&\frac{4}{7}\sqrt{\frac{5\pi}{6}}\eta x\left[1+x\left(-\frac{4075}{4032}+\frac{67}{48}\eta\right)\right.\\
    &\left.+x^{3/2}\left(\frac{-2813+756\eta}{2400}\chi_S+\frac{3187}{2400}\delta\chi_A\right)\right],
\end{split}
\label{PNexpr}
\end{equation}
where $x$ is the frequency variable $x=\left(\frac{d\phi}{dt}\right)^{2/3}$, $\eta=\frac{m_1m_2}{(m_1+m_2)^2}$, $\chi_S=\frac{m_1\chi_1+m_2\chi_2}{m_1+m_2}$, $\chi_A=\frac{m_1\chi_1-m_2\chi_2}{m_1+m_2}$ and $\delta=\sqrt{1-4\eta}$. We obtain the phase evolution using the $(2, 2)$ mode of the {\tt IMRPhenomXHM} model for each combination of parameters.\\

In order to fit this expression to the NR data, we add a correction in terms of $x$. We have to take into account that the PN expression loses accuracy as we approach the time of the merger ($t=0$), therefore we compute the residuals between the PN expression and the NR waveforms up to $t=-20M$:
\begin{equation}
    \text{Res}(x)=h_{2,0}^{\text{PN}}-h_{2,0}^{\text{NR}}.
\end{equation}
We then fit the residuals with the following expression, which corresponds to higher orders in $x$ with fitting parameters $\alpha$ and $\beta$, and an off-set parametrized by $c$:
\begin{equation}
    c+\alpha x^3+\beta x^4.
   \label{corrinsp} 
\end{equation}
The coefficient $c$ is used to fix the constant of integration to be consistent with the NR simulations as they start at a time different from $-\infty$. So we add this freedom to be able to better fit to the data, but for low frequencies, i.e. $x\rightarrow 0$, the amplitude $R/M h_{2,0}^{\text{(PN,mem)}}$ must tend to zero too. Therefore, in the final waveform we subtract this constant off-set in order to recover this behavior which is already satisfied by the PN expression. For the purpose of comparison with NR, one can always fix this freedom by shifting the waveforms. Regarding the \textit{Ansatz} proposed, we also tried to add higher powers of $x$ and to compute the residuals up to a different end-time, but since the results of the coefficients did not change significantly, we decided to extend this \textit{Ansatz} up to $t=-20M$ and to use the expression in (\ref{corrinsp}) so the complexity of the \textit{Ansatz} is not increased unnecessarily with higher powers.

Here we are only interested in the behavior of the waveforms in the inspiral regime, where the oscillatory part of this mode is negligible. So, regarding the dataset, we use the SXS simulations for which we compute the memory contribution using the {\tt sxs.waveforms.memory.add\_memory} option, which includes in the calculation all the available higher modes in the corresponding SXS simulation \cite{Mitman:2020bjf}. We do not use the EMRI simulations for the inspiral, because getting their evolution in the inspiral regime is computationally expensive. Instead, for high values of $q$, we assume that the amplitude of the memory mode must tend to zero in the extreme mass ratio limit. Therefore, for $\eta=0$ we simply set $c=\alpha=\beta=0$, and we check that the behavior that we get is consistent with the surfaces across the parameter space obtained using the SXS simulations. We fit the coefficients $c$, $\alpha$ and $\beta$ across the three-dimensional parameter space defined by the symmetric mass ratio $\left(\eta=\frac{m_1m_2}{(m_1+m_2)^2}\right)$, the effective spin $\left(S=\frac{m^2_1\chi_1+m^2_2\chi_2}{m^2_1+m^2_2}\right)$ and the spin difference ($\Delta\chi=\chi_1-\chi_2$), following a hierarchical procedure to construct a parameterized fit that avoids overfitting as described and implemented in \cite{Jimenez-Forteza:2016oae}. We first perform two one-dimensional fits: one for the nonspinning data and another one for the equal mass, equal spin simulations. Then these two fits are extended to a two-dimensional fit as a function of the symmetric mass ratio and the effective spin. Finally, the dependence on the spin difference is added as a correction to the previous fit by computing the dependence of the data on this parameter for fixed values of the symmetric mass ratio.

Once we have analytical expressions for $c(\eta,S,\Delta\chi)$, $\alpha(\eta,S,\Delta\chi)$ and $\beta(\eta,S,\Delta\chi)$, we are able to reconstruct the inspiral part of the waveform by adding the correction in (\ref{corrinsp}) to the PN expression in (\ref{PNexpr}) for any combination of parameters. We observe that for $t\rightarrow -\infty$, the strain decays following a power law of the type $A |t-t_0|^{-\gamma}$ with $t_0$ the time of the merger and $\gamma\approx1/4$ at the lowest order.

\subsubsection{Late inspiral, merger and ringdown}

\begin{figure*}[htp]
    \centering
    \includegraphics[width=0.9\textwidth]{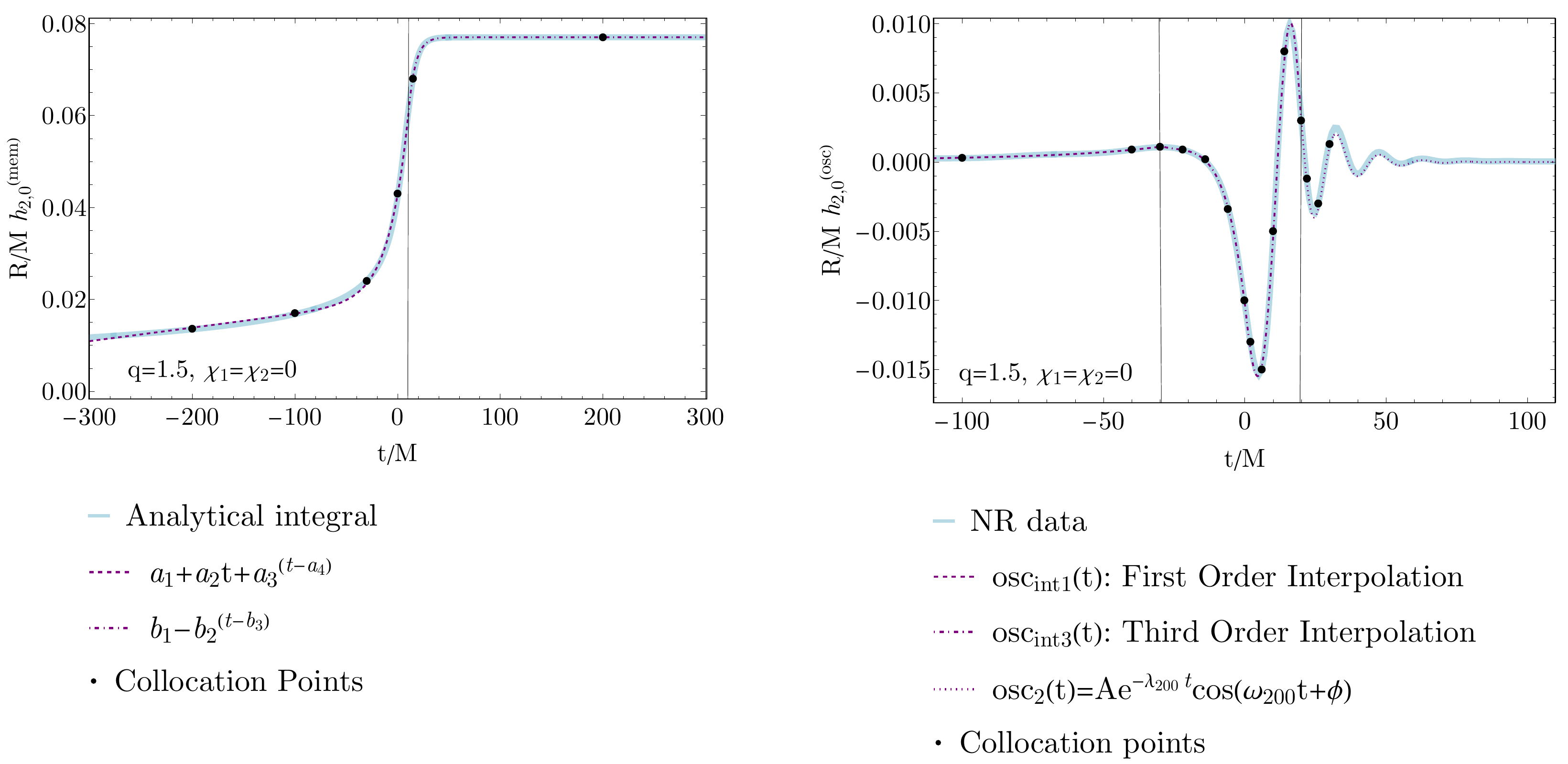}
    \caption{Left: fits of the analytical ansätze to the analytical integral of the memory contribution of the $(2,0)$ mode. The black dashed line indicates the time $t=10M$ that delimits the two time regions (\ref{ansatzmem}). Right: fits to the NR curve of the oscillatory contribution of the $(2,0)$ mode. The black dashed lines indicate the times $t=-30M$ and $t_2$ that delimit the three different time regions. In both curves, the black points indicate the collocation points that are fitted across the parameter space. These curves are for a nonspinning BBH system with $q=1.5$, corresponding to the NR simulation SXS:BBH:0194 in the case of the oscillatory part.}
    \label{fig:curvefits}
\end{figure*}

Finding an analytical expression which reproduces the behavior over the full range of times that covers the late inspiral, the merger and the ringdown turns out to be challenging. Therefore, we split it into two different regions. For each of these regions we propose a different \textit{Ansatz}, using 7 free coefficients ($a_{1,2,3,4}$ and $b_{1,2,3}$):
 \begin{equation}
\left\{
\begin{split}
    &\text{mem}_1(t)=a_1+a_2 t+a_3^{(t-a_4)} \quad \text{if}\quad -300M\leq t\leq 10M,\\
    &\text{mem}_2(t)=b_1-b_2^{(t-b_3)} \quad \text{if}\quad 10M< t\leq 300M.
\end{split}
\right.
\label{ansatzmem}
\end{equation}
We found that this simple choice of expressions is able to approximate the phenomenology of the memory waveform sufficiently well. 
 
We fit both expressions in (\ref{ansatzmem}) to the waveforms in order to obtain the value of the coefficients. As we have done for the inspiral regime, we try to fit this set of 7 coefficients across the parameter space, but we find that with this procedure we do not recover the waveforms very accurately. For this reason, instead of directly fitting the coefficients, we represent a waveform by its values at a set of 6 time values, which we refer to as collocation points. For each waveform in our dataset we thus first fit our \textit{Ansatz} and determine the 7 coefficients, and then we determine from the resulting waveform a dataset of 6 pairs $\left(t/M,\; R/M h_{2,0}^{\text{(mem)}}\right)$. The collocation points are placed at the  times $t = (-200M, -100M, -30M, 0, 15M, 200M)$. We then fit the values at the collocation points across parameter space. Fitting collocation points instead of \textit{Ansatz} coefficients has previously been found to yield more robust parameter space fits   in the construction of other phenomenological models such as {\tt IMRPhenomD} \cite{Khan:2015jqa}, {\tt IMRPhenomX} \cite{Pratten:2020fqn} and {\tt IMRPhenomT} \cite{Estelles:2020osj}.

Then, for each combination of parameters in the dataset, we use the analytical expressions of the collocation points as a function of $\eta$, $S$ and $\Delta\chi$ and the ansätze in (\ref{ansatzmem}) to solve numerically the system of equations for the coefficients. For instance,
\begin{equation}
    a_1+a_2 t+a_3^{(t-a_4)}=R/M h_{2,0}^{(\text{mem})}(t).
\end{equation}

We use one less collocation point than the number of coefficients because in the second region we also use the collocation point at $t=0$ so that both regions share information and therefore, they are consistent. Hence, we have four equations using the collocation points at $t=(-200M, -100M, -30M, 0)$ to obtain $a_{1,2,3,4}$ and three equations using $t=(0, 15M, 200M)$ to obtain  $b_{1,2,3}$. Once we have determined all the coefficients, we are able to reconstruct the waveform for any combination of parameters. Combining the results for the inspiral regime up to $t=-30M$ and the results for the late-inspiral, merger ringdown from $t=-30M$ we have the full description of the memory contribution of the waveform.

In the left panel of Fig. \ref{fig:curvefits} we present an example of a fit for the memory part corresponding to the parameters $q=1.5\;\chi_1=\chi_2=0$. The purple dashed and dash-dotted curves are the analytical ansätze fitted to the light blue curve and the black points are the collocation points that we fit. The black vertical line delimits the two different regions at $t=10M$.

\subsection{Fits for the oscillatory contribution}
\label{subsec:fits_osc}

For the fits of the oscillatory contribution we use the $(2,0)$ mode from the SXS Catalog and the EMRI simulations. We characterize the curves by determining the turning points $\left( \Ddot{h}_{2,0}^{\text{(osc)}}=0\right)$. Then, we use this data to split the curves into three different regions: the first one up to $t=-30M$, the second one from $t=-30M$ up to the fourth turning point, denoted by $t_2$ and the last one from $t_2$ up to $t=300M$. We choose $t=-30M$ to be the time at which approximately the inspiral regime ends and so the behavior of the curves changes to reproduce the oscillations in the merger-ringdown. On the other hand, we extend it up to $t=300M$ because, in all cases, at this time the system stabilizes so that the curves are constant and have a value very close to or equal to zero.\\
For the last region, i.e. $t_2\leq t\leq300M$, we propose an analytical \textit{Ansatz} which depend on two coefficients ($A$ and $\phi$) and we use quasinormal mode (QNM) frequencies \cite{QNMs} to describe the ringdown oscillations ($\omega_{200}$ and $\lambda_{200}$):
\begin{equation}
     \text{osc}_2(t)=Ae^{-\lambda_{200}t}\cos(\omega_{200} t+\phi)\quad \text{if}\quad t> t_2.
\label{ansatzosc}
\end{equation}
We fit this \textit{Ansatz} to the NR data so that we can obtain the values of these coefficients and also of the time variable $t_2$, across the parameter space.

For the first two time regions, i.e. $t\leq t_2$, similar to the memory contribution, we utilize  collocation points to represent the waveform in the parameter space fits, and we place these collocation points at  times $t = (-30M, -22M, -14M, -6M, 0, 2M, 6M, 10M, 14M, 18M, 22M,$ $26M, 30M)$. Again, we fit the values of $R/M h_{2,0}^{\text{(osc)}}$ at these collocation points across the parameter space with the method described in \cite{Jimenez-Forteza:2016oae}.\\
Once we have the expressions that describe the behavior of these quantities as a function of $\eta,\; S$ and $\Delta\chi$, we are able to reconstruct the waveform for any combination of these parameters. For the first region, $t\leq-30M$, we perform a first order interpolation of the collocation points, whereas for the other region ($-30M\leq t\leq t_2$), the interpolation is of third order. These two parts are labeled as osc$_{\text{int1}}$ and osc$_{\text{int3}}$, respectively.\\
In the right panel of Fig. \ref{fig:curvefits} we present the oscillatory contribution of the NR waveform SXS:BBH:0194 together with the curves that are fitted represented by the purple dashed, dash-dotted and dotted lines, respectively. The black vertical lines delimit the three time regions. The black points indicate the position of the collocation points which are fitted across the parameter space.\\
With all the time regions characterized, we join them and we have the full picture of the oscillatory contribution. At this stage we have both contributions of the waveform ready to be added together so we can reconstruct the waveform of the full $(2,0)$ spherical harmonic.

\subsection{Reconstruction of the full waveform}
\label{subsec:fullWF}
The final model can be summarized as follows:

\begin{itemize} 

\item In the region 1, $t\leq -30M$, we have the analytical expression from the PN approximation plus the correction and the first order interpolation of the oscillatory contribution: $R/M h_{2,0}^{\text{(PN,mem)}}(x)+c+\alpha x^3+\beta x^4+\text{osc}_{\text{int1}}(t)$.

\item In the region 2, $-30M\leq t\leq 10M$, we sum the third order interpolation of the oscillatory part and the first region of the memory contribution: $\text{osc}_{\text{int3}}(t)+\text{mem}_1(t)$.

\item In the region 3, $10M\leq t\leq t_2$, we sum the third order interpolation of the oscillatory part and the second region of the memory contribution: $\text{osc}_{\text{int3}}(t)+\text{mem}_2(t)$.

\item In the region 4, $t_2\leq t\leq 300M$, we sum the the analytic \textit{Ansatz} of thoscillatory part and the second region of the memory contribution: $\text{osc}_2(t)+\text{mem}_2(t)$.

\end{itemize}

The four different regions must be joined together in a continuous waveform. To do so, we impose continuity between the regions using three constants that need to be determined using the boundary conditions: $\tau, \mu$ and $\nu$. We obtain the value of $\tau$ using the condition at $t=-30M$:
\begin{equation}
\begin{split}
 &\text{(PN+correction)}(t=-30M)+\text{osc}_{\text{int1}}(t=-30M)\\
    &=\tau\left[\text{osc}_{\text{int3}}(t=-30M)+\text{mem}_1(t=-30M)\right].
\end{split}
\end{equation}
Once $\tau$ is known, we use it to determine the values of $\mu$ and $\nu$ with the two remaining conditions at $t=10M$ and $t=t_2$:
\begin{equation}
\begin{split}
    &\tau\left[\text{osc}_{\text{int3}}(t=10M)+\text{mem}_1(t=10M)\right]\\
    &=\mu+\nu\left[\text{osc}_{\text{int3}}(t=10M)+\text{mem}_2(t=10M)\right],
\end{split}
\end{equation}
\begin{equation}
\begin{split}
    &\mu+\nu\left[\text{osc}_{\text{int3}}(t=t_2)+\text{mem}_2(t=t_2)\right]\\
    &=\text{osc}_2(t=t_2)+\text{mem}_2(t=t_2).
\end{split}
\end{equation}
The waveforms are $C^0$, since making them smooth would introduce additional functional dependencies, increasing the complexity of the analytical expressions. This approach is consistent with the strategy used in the rest of {\tt IMRPhenom} models. At the current level of accuracy, $C^0$ waveforms are sufficient also considering that the (2,0) is a subdominant mode.

\section{Accuracy of the model}
\label{sec:accuracy}

In order to check how the model performs qualitatively for few different combinations of parameters, we demonstrate, in Fig. \ref{fig:multiplemodel}, four cases comparing the NR waveform (light blue curve) and the waveform reconstructed using our model (purple dash-dotted curve) together with the time evolution of the residuals between them (yellow dashed curve). We choose different combinations of the mass ratio and spins in order to compare the accuracy in various regions of the parameter space.

\begin{widetext}
\begin{center}
\begin{figure}[htp]
\includegraphics[width=0.8\textwidth]{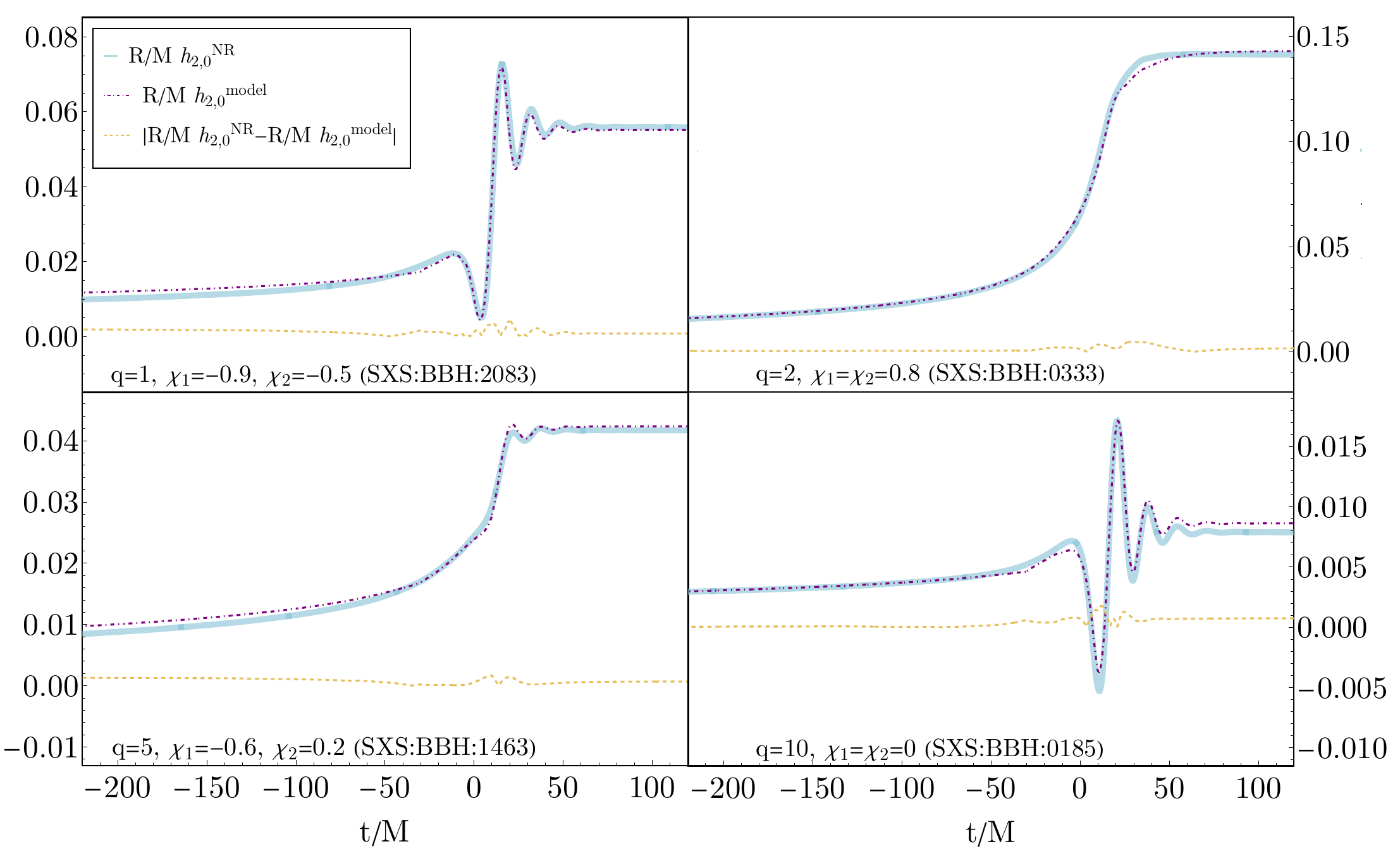}
    \caption{Comparison between the waveforms reconstructed from our model and the NR data for four different combinations of parameters indicated in the plots together with the corresponding SXS simulation. The legend within the upper left plot is common to all of them. The yellow dashed curves are the time evolution of the residuals between the two waveforms. For the visual comparison the model waveforms are shifted to match the off-set at zero frequency of the numerical waveforms, therefore $h_{2,0}(f=0)\neq0$.
    \label{fig:multiplemodel}
    }
\end{figure}
\end{center}
\end{widetext}

To quantitatively estimate the accuracy of the model, we compare the waveform that we obtain through reconstruction with the corresponding NR waveform from the SXS Catalog through the calculation of the mismatch \cite{Bruegmann:2007bri}. To this end, we first introduce the usual definition of the inner product between two waveforms (denoted by $h$ and $g$) weighted by the noise estimated from the detector's PSD,
\begin{equation}
\label{innerprod}
    \langle h|g\rangle\equiv 4 \text{Re}\left[\int_{f_{\text{min}}}^{f_{\text{max}}}\frac{\Tilde{h}(f)\Tilde{g}^*(f)}{S_n(f)}df\right].
\end{equation}
Then the match between these two waveforms is defined as
\begin{equation}
    \mathcal{M}\equiv\underset{\Delta t, \Delta \varphi}{\max}\frac{\langle h|g\rangle}{\sqrt{\langle h|h\rangle \langle g|g\rangle}},
\end{equation}
where overlap is maximized over the relative time shift  $\Delta t$ and phase shift $ \Delta \varphi$. To compare the different waveforms we will use the mismatch, which is defined as $1-\mathcal{M}$. Therefore, as two waveforms become more similar, the value of the mismatch tends to zero.

As indicated in (\ref{innerprod}), we need to compute the Fourier transform of the waveforms. When considering the full picture of the $(2,0)$ spherical harmonic, this calculation is not straightforward. As we have seen, when including the memory growth, we get a nonperiodic signal, and therefore performing the Fourier transform on this signal directly leads to the introduction of artifacts that are not physical. In order to avoid this, before transforming the signal into the frequency domain, we modify the waveform so that we get a periodic behavior. First we pad the waveform at the end with a constant padding. We do this for different lengths of time and we check that the results do not change for the different lengths. Then, we apply a Planck window \cite{McKechan:2010kp} over both ends of the waveform to make it periodic, so it goes to zero for $t\rightarrow -\infty$ as well as for $t\rightarrow \infty$. Finally, we apply a zero-padding at both ends. Once we have periodic waveforms, we bring them to the frequency domain performing a discrete Fourier transform. In a future publication we discuss a refined method for performing the Fourier transform~\cite{Valencia:2024zhi}.

\begin{figure}[htp]
\includegraphics[width=0.9\columnwidth]{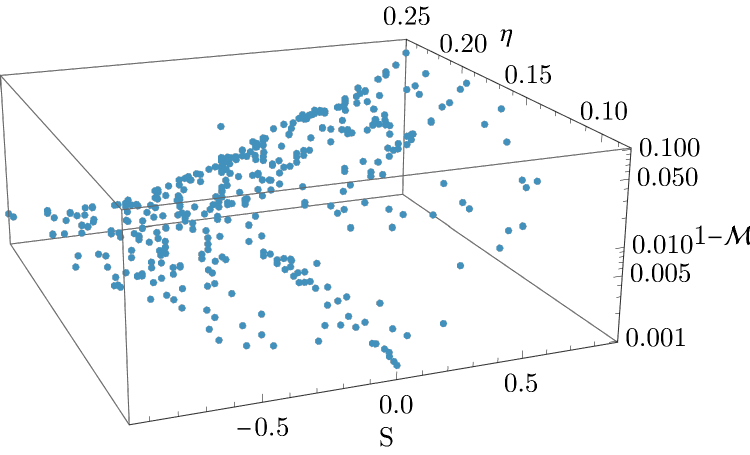}
    \caption{Mismatch of the (2,0) mode between our model and the SXS simulations across the parameter space defined by $\eta$ and $S$ for the Advanced LIGO detector and $M=100M_{\odot}$.
    \label{fig:matchpar}
    }
\end{figure}

The calculations carried out here are performed taking into account the PSD of the Advanced LIGO detector \cite{Sensitivitycurves} and a total mass of the system of $100M_{\odot}$. We compute the mismatch between the 431 simulations from the SXS Catalog and the waveform obtained with our model with the corresponding parameters and we present the results for the (2,0) mode in Fig. \ref{fig:matchpar}, where we can check the accuracy across the parameter space defined by the symmetric mass ratio and the spin variable.

We see that the mismatch increases as the value of the variable $S$ grows, therefore we can conclude that our model becomes less accurate as the spins take high positive values. In this region, the highest values of mismatch are obtained for the values of $\eta\in(0.15,0.19)$. In particular, the less consistent waveforms are the ones corresponding to the simulations SXS:BBH:0293 ($q=3.0,\; \chi_1=\chi_2=0.85$), SXS:BBH:1434 ($q=4.4,\; \chi_1=\chi_2=0.80$) and SXS:BBH:1452 ($q=3.6,\; \chi_1=0.80,\;\chi_2=-0.43$). For these parameters, the memory amplitude is more important than the ringdown oscillations, hence the full waveform of the mode is dominated by the growth in amplitude due to the memory. See the right top panel of Fig.~\ref{fig:multiplemodel} for an example. In these cases, our model presents some slight oscillations which are not consistent with the SXS waveforms. We checked that the most important source of inaccuracy in the model comes from the memory contribution. Regarding time domain comparisons, this behavior was expected since the direct fits of the ansätze to the data were slightly worse for this part of the waveforms, due to its phenomenology, than for the oscillatory part of the waveform, which is very well recovered across all the parameter space. We also have to take into account that the oscillatory part is directly calibrated to the NR data, whereas the memory part is calibrated using the BMS integral of the modes from the {\tt IMRPhenomXHM} phenomenological model, which can lead to a first loss of accuracy with respect to NR.

\begin{figure}[htp]
\includegraphics[width=1\columnwidth]{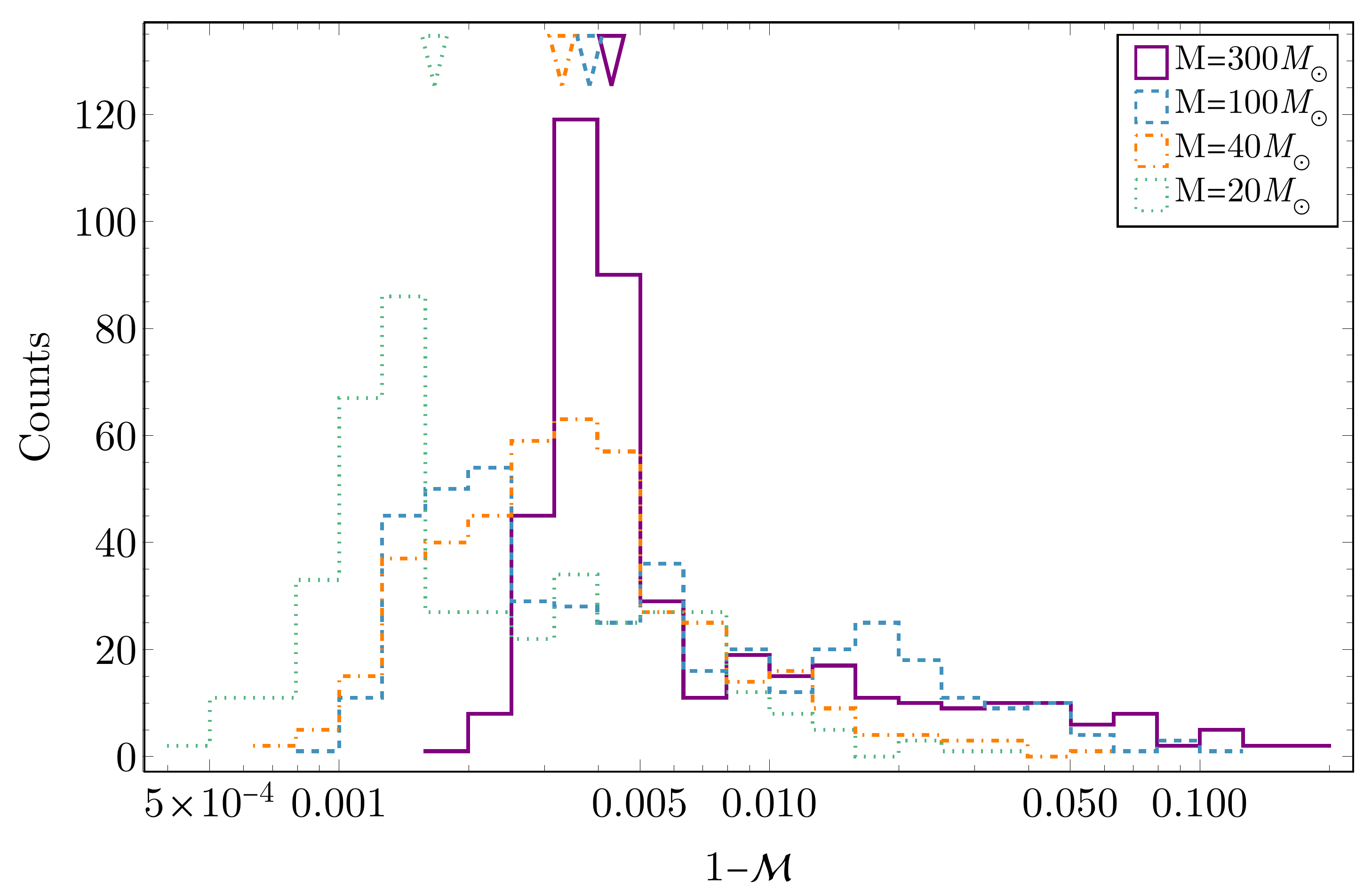}
    \caption{Distribution of mismatches of the (2,0) mode between our model and the 431 SXS simulations for different values of the total mass, indicated in different colors for the Advanced LIGO detector. The vertical dashed lines indicate the median value of each set of data.
    \label{fig:hist_masses}
    }
\end{figure}

Fig. \ref{fig:matchpar} corresponds to a single value of the total mass of the system, but we studied the mismatches for four different values of the total mass, $M=20,40,100,300M_{\odot}$ and we present the histograms of the results in Fig. \ref{fig:hist_masses}. The triangles in the plot indicate the median values of each dataset (matching the line style and color), from where we can see that, as the mass increases, the mismatch also does, e.g. for $M=20M_{\odot}$, we have a median value of 0.0017, whereas for $M=300M_{\odot}$, the median is 0.0043.

\begin{figure}[htp]
     \centering
     \begin{subfigure}
         \centering
         \includegraphics[width=1\columnwidth]{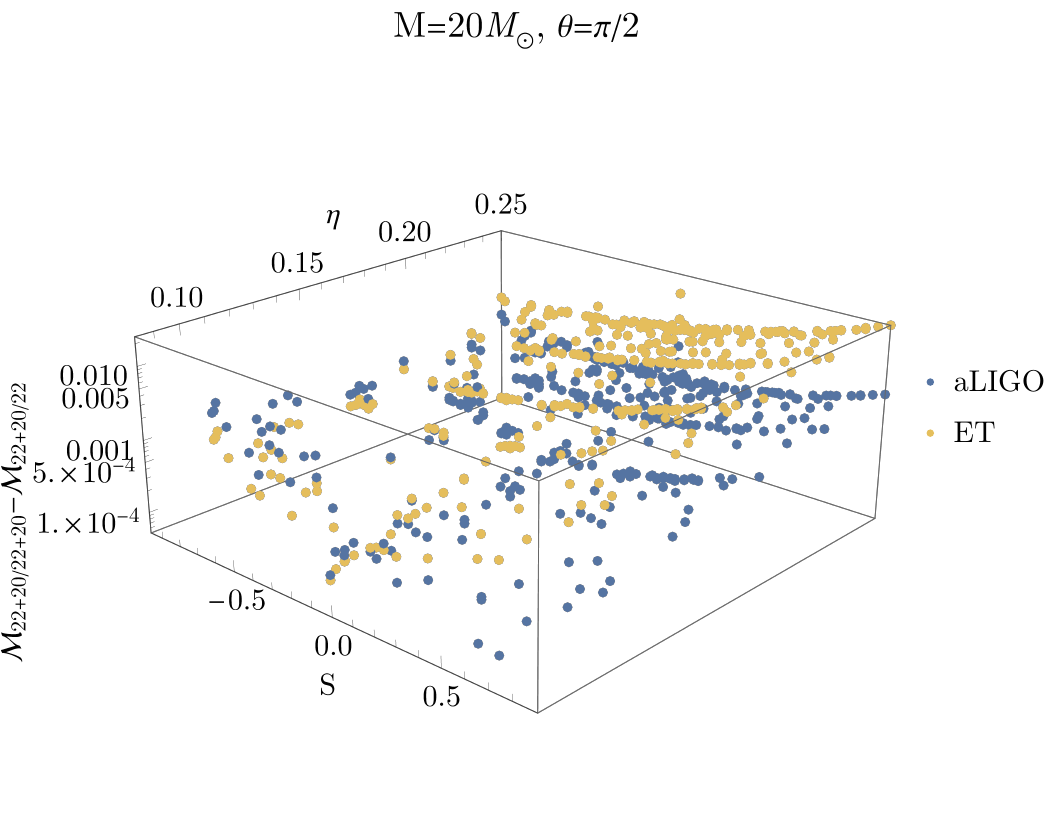}
     \end{subfigure}
     \hfill
     \begin{subfigure}
         \centering
         \includegraphics[width=1\columnwidth]{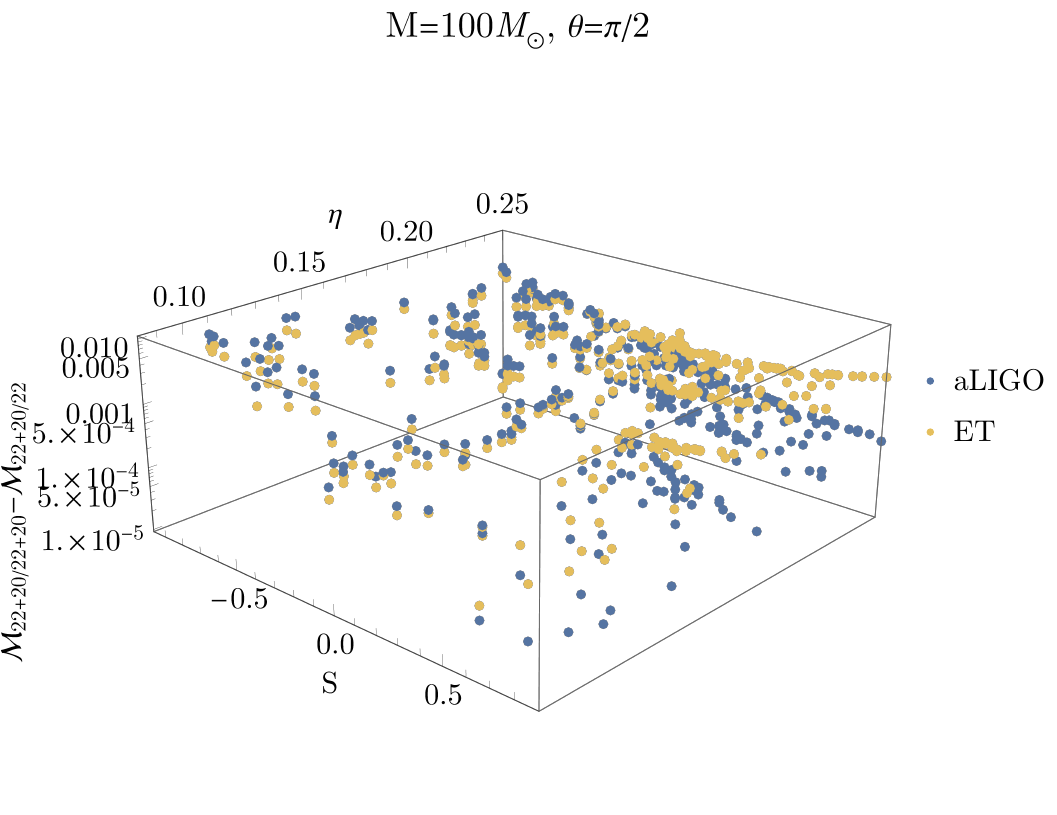}
     \end{subfigure}
        \caption{Dependence on the symmetric mass ratio and the effective spin of the match difference between using and not using the (2,0) mode in the model waveform (denoted as $\mathcal{M}_{22+20/22+20}$ and $\mathcal{M}_{22+20/22}$ respectively) when the target waveform includes both the $(2,\pm2)$ and the (2,0) modes. Each plot represents a different value of the total mass and each color a different detector, where aLIGO is the abbreviation for Advanced LIGO and ET for Einstein Telescope.}
        \label{fig:M22_20masses}
\end{figure}

In order to test our model, we also need to perform match calculations with different mode content. We have already computed the match between the $(2,0)$ mode of the model against the calibration dataset. Now we compare the match we achieve when we take as the target the waveform containing both the $(2,0)$ and $(2,2)$ modes, and for the template either only the  $(2,2)$ mode, or both modes. We contrast this with the match of only $(2,2)$ modes of the target and template waveforms. Our model will be beneficial for data analysis when adding the $(2,0)$ to the $(2,2)$ mode improves the match with a target waveform that contains both modes. Note that when only using the $(2,0)$ mode we do not have to optimize over the phase, and when only using the $(2,2)$ mode we can perform this optimization analytically. When using both modes we use a numerical optimization. In Fig. \ref{fig:M22_20masses} we present the difference in the match when changing the mode content of the template, where $\mathcal{M}_{22+20/22+20}$ indicates the match between the SXS and the model waveforms both containing the $(2,\pm2)$ and the (2,0) modes, whereas $\mathcal{M}_{22+20/22}$ represents the match when we remove the (2,0) mode from the model waveform. We can see the dependence over the parameter space defined by the symmetric mass ratio and the effective spin, in the top plot for a total mass of the system of 20$M_{\odot}$ and in the bottom plot for 100$M_{\odot}$. In both cases we choose an inclination angle of $\pi/2$ since this value is the one giving the highest contribution of the (2,0) due to its dependence on $\theta$ through $^{-2}Y_{2,0}(\theta)$. The blue dataset represents the results for the Advanced LIGO detector and the orange dataset, for the Einstein Telescope (ET) , for which we used again the PSDs in \cite{Sensitivitycurves}.

The difference in the matches is always positive, which means that adding the (2,0) mode improves the accuracy of the model, as one could expect. Since the $(2,\pm2)$ modes are already very accurate and the (2,0) is some orders of magnitude smaller than the dominant modes, the maximum improvement in the match that we get is of  $\sim0.025$. As we approach the equal mass region, the match difference increases as the memory effect is more prominent. Since the memory appears as a low frequency component, for the values of the spins for which the memory amplitude is larger (high positive), the improvement is larger for ET than for aLIGO for both values of the mass since ET has a better sensitivity, especially at low frequencies.

\subsection{Comparison with other models}
\label{subsec:compmodels}

\begin{figure}[htp]
\includegraphics[width=0.87\columnwidth]{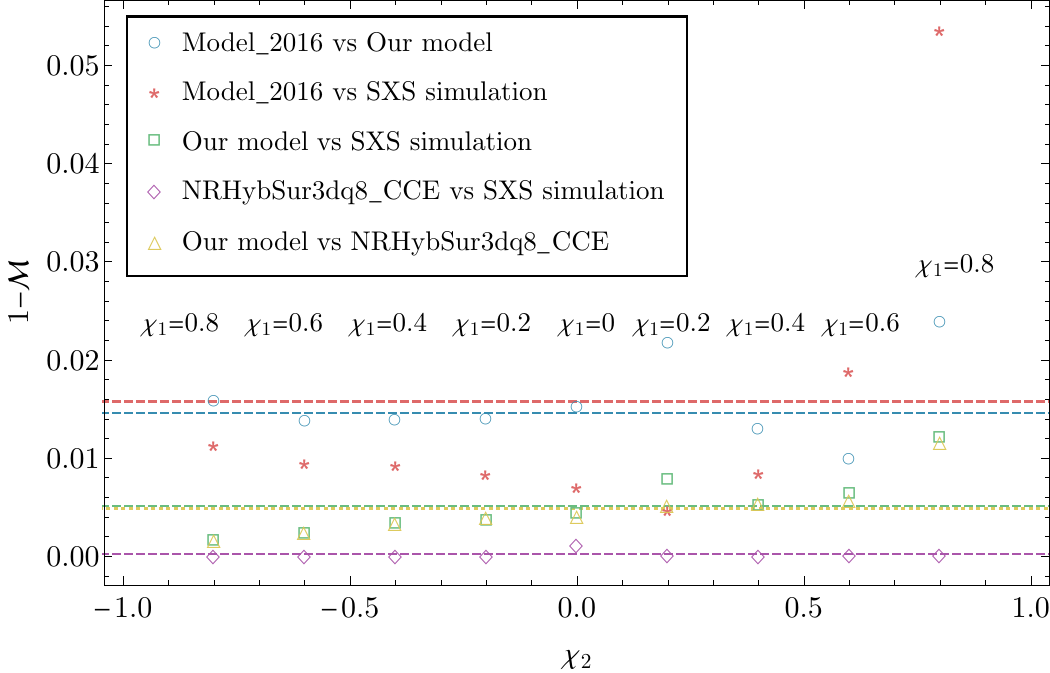}
    \caption{Mismatches between our model, the model from \cite{Cao_2016}, the model from \cite{Yoo:2023spi} and the corresponding SXS simulations. From left to right: SXS:BBH:0327, SXS:BBH\_ExtCCE:0007, SXS:BBH\_ExtCCE:0006, SXS:BBH\_ExtCCE:0005, SXS:BBH:0001, SXS:BBH:0150, SXS:BBH\_ExtCCE:0003, SXS:BBH:0152, SXS:BBH:0155. All the configurations are for $q=1$. The horizontal dashed lines indicate the mean value of the mismatch for each set of points. All the quantities are computed for the Advanced LIGO detector using a total mass of $100M_{\odot}$. For $q=1$, we have $S=\chi_S=\frac{\chi_1+\chi_2}{2}$.
    \label{fig:MmodelsSXS}
    }
\end{figure}

In this section we analyze and compare the accuracy of the model presented in this work with the model introduced in 2016 \cite{Cao_2016} and the {\tt NRHybSur3dq8\_CCE} \cite{Yoo:2023spi} using the mismatch calculations only for the (2,0) mode. We compare the three models by computing the mismatch between them and we compare their accuracy by computing the mismatch between each of the models and the corresponding SXS simulation. For the combinations of spin components for which the SXS simulations plus the memory correction is not available, we use the Ext-CCE waveforms from the SXS Catalog \cite{SXS:ExtCCE}. The mismatch dependence on the individual spin components is presented in Fig. \ref{fig:MmodelsSXS}, where the values of the first component are given within the plot and the second component is represented in the horizontal axis.  The calculations are performed for combinations of parameters for which the 2016's model can be evaluated and for which the corresponding NR simulations are available ($q=1,\; \chi_S=\{0,0.2,0.4,0.6,0.8\}$). Note that the first component is positive for all the cases.

For all the sets of points except the purple diamonds, we get a similar behavior, with the mismatch increasing as the effective spin does as well. Regarding the comparison between the 2016's model and our model (blue circles), the mismatch increase when both spin components are positive and this is due to the fact that the waveforms for these parameters generated with the previous model show oscillations in the merger-ringdown regime which are not present in the NR simulations and neither in our model. This is also the reason why the accuracy of this model against the SXS data (red stars) is worse. Comparing the mismatch results between our model and the SXS simulations (green squares) with the blue circles, we can conclude that our model gives a better accuracy than the previous one in all the cases, but its performance could be improved for $\chi_S\geq0.6$. This difference can also be noticed through the comparison of the mean values of each set of points (dashed lines in the plot). We check that the mean mismatch between the previous model and both our model and the SXS simulations leads to a very similar value. In contrast, the consistency between our model and the SXS data is much better as the mean mismatch is smaller by a factor of three. Now we focus on the comparison with the {\tt NRHybSur3dq8\_CCE} model (purple diamonds). For values of the spins of $\chi_S\leq0$, the mismatches of our model are comparable, but for positive values of the spin components, the mismatches for our model increase, whereas for the surrogate remain at values near zero. In any case, it is expected that this NRsurrogate model is more accurate as it is constructed hybridizing PN and EOB (effective-one-body) waveforms and performing a BMS frame fixing, in order to ensure that all the waveforms are in the same frame (see \cite{Mitman:2022kwt,Mitman:2024uss} for details). In our case, we were not able to carry out this procedure of frame fixing since, to compute the BMS charges, one needs to have the whole set of Weyl scalars ($\Psi_0,\Psi_1,\Psi_2,\Psi_3,\Psi_4$), but in the public SXS Catalog \cite{SXS:catalog} that we use, the only available quantities are the strain $h$ and the $\Psi_4$ scalar.

In Appendix \ref{appendix:compmodel} one can find a qualitative comparison of the waveforms that are used in the calculations of the mismatch previously presented.

\section{Signal-to-noise ratio}
\label{sec:SNR}

Finally, we perform some calculations of the signal-to-noise ratio (SNR) in order to investigate how including memory in the signal affects to the detectability of the $(2,0)$ mode. In gravitational wave data analysis, the SNR is used to determine if an event can be measured in a detector. Considering a signal $h(t)$, and the definition of the inner product previously introduced in (\ref{innerprod}), the optimal SNR is computed in the frequency domain as
\begin{equation}
    \rho\equiv\sqrt{\langle h|h\rangle}=\sqrt{4\int_0^{\infty}\frac{|\Tilde{h}(f)|^2}{S_n(f)}df}.
\end{equation}

We perform some calculations of the SNR using the waveforms obtained with our model, both including and not including the memory contribution. In this way, we can check how both the memory and oscillatory contributions affect the detectability of the $(2,0)$ mode. In all the calculations we use the PSD of the Advanced LIGO detector \cite{Sensitivitycurves}.

\begin{figure}[htp]
\includegraphics[width=0.9\columnwidth]{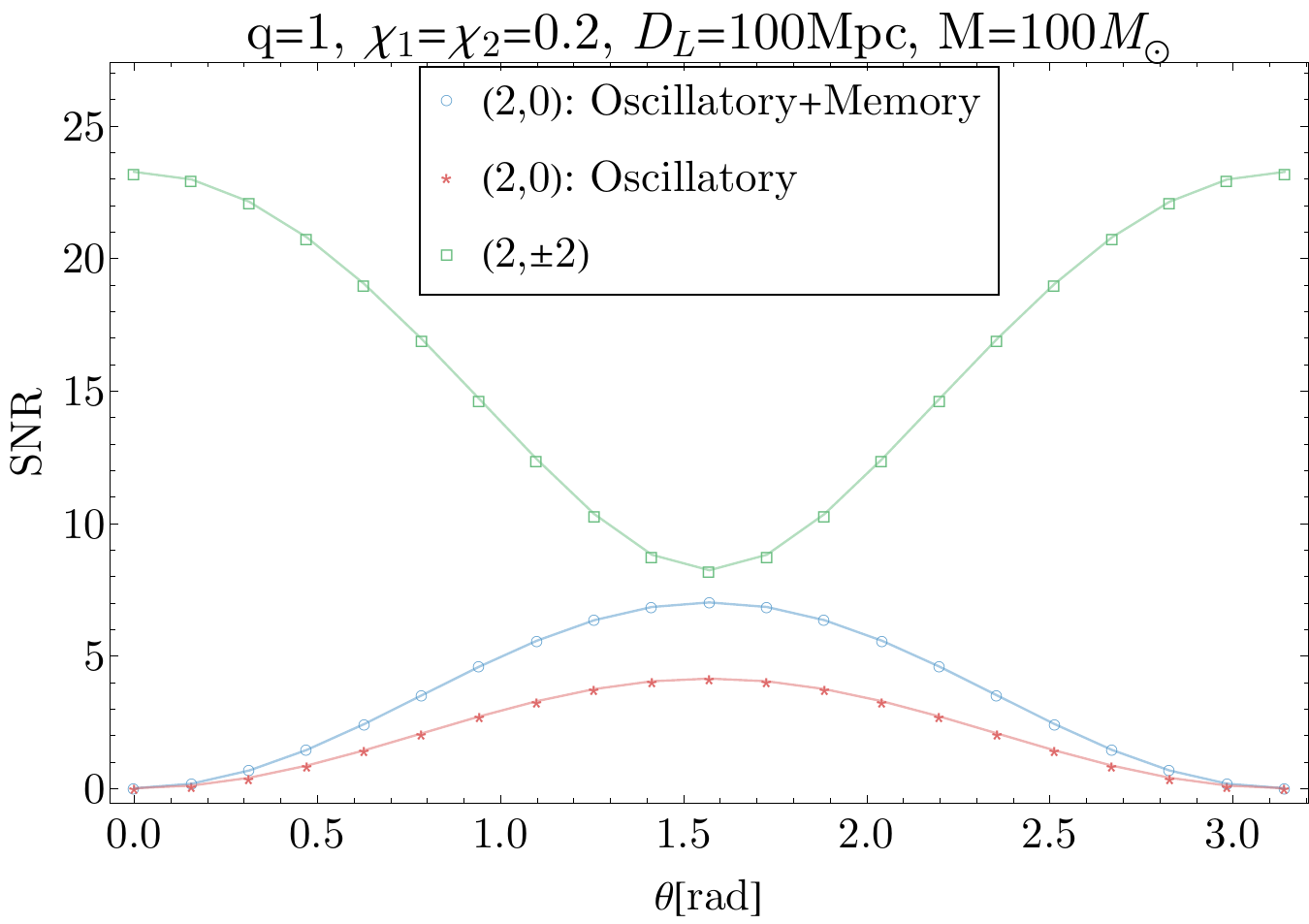}
    \caption{Dependence of the SNR on the inclination of the binary for an equal mass, equal spin system with the parameters stated in the figure. Comparison between the SNR of the dominant $(2,\pm2)$ mode (from the simulation SXS:BBH:0150) and the $(2,0)$ mode with and without the memory contribution computed with our model for the Advanced LIGO detector.
    \label{fig:SNRinc}
    }
\end{figure}

First we analyze the different dependence that this mode and the dominant modes have on the inclination angle. In Fig.~\ref{fig:SNRinc} one can check that the SNR for the $(2,\pm2)$ is maximum for face-on systems, the case for which the $(2,0)$ mode vanishes. In contrast, the SNR for the $(2,0)$ is maximum for edge-on systems, when the SNR of the dominant modes is minimum. Therefore, the events mostly detected will correspond to face-on systems, the case when the $(2,0)$ vanishes. In the situation where we can detect both modes, this information can be used to break the distance-inclination degeneracy. See \cite{xu2024enhancing} for a comprehensive study on this topic.

Focusing on the blue and red curves, we notice that the waveform with memory slightly improves the SNR, especially for values of $\theta$ around $\pi/2$. The maximum value of the blue curve reaches SNR = 7.01, whereas the red curve only gets to SNR = 4.14. Hence, for this specific combination of parameters in an optimally-oriented system, the improvement is of a $41\%$ with respect to the oscillatory waveform's SNR. Therefore, the main contribution to the SNR of the $(2,0)$ mode comes from the oscillatory component as the memory only adds a low-frequency contribution, but it could be useful to reach the detection threshold for this mode.

\begin{figure}[htp]
\includegraphics[width=0.85\columnwidth]{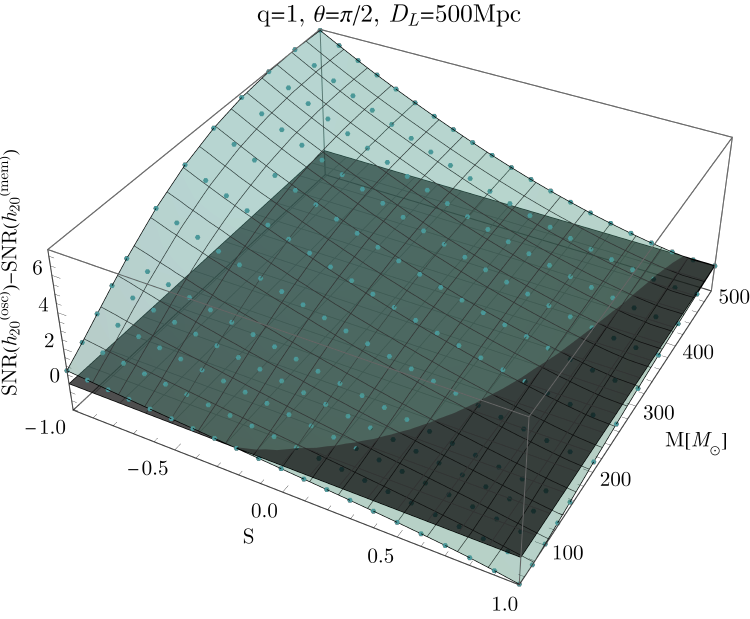}
    \caption{Dependence on the total mass and the effective spin of the difference between the SNR of the oscillatory contribution and the memory contribution of the (2,0) mode, for an equal mass, edge-on system at a distance of 500 Mpc for the Advanced LIGO detector. We consider only equal spin systems. The black surface corresponds to the case of SNR$\left(h_{2,0}^{\text{(osc)}}\right)$=SNR$\left(h_{2,0}^{\text{(mem)}}\right)$.
    \label{fig:SNRmassinc}
    }
\end{figure}

Next, we check the joint dependence of the SNR on the total mass and the effective spin of the binary system. We present in Fig. \ref{fig:SNRmassinc} the difference between the SNR of both contributions of the (2,0) mode in order to see the relation between these two quantities for these combinations of parameters. We add a surface for which SNR$\left(h_{2,0}^{\text{(osc)}}\right)$=SNR$\left(h_{2,0}^{\text{(mem)}}\right)$ in order to clearly see where one contribution gives a larger SNR than the other. We choose the inclination to be edge-on since we know from Fig. \ref{fig:SNRinc} that this is the optimal case for which the memory is favored. For the same reason we choose an equal mass system.

For a large part of the parameter space, we see that the blue surface is above the black one, meaning that the SNR of the oscillatory contribution dominates over the SNR of the memory. When the effective spin becomes high positive, the memory contribution overtakes the ringdown oscillations in the waveform, therefore giving a higher contribution to the SNR. This happens for lower values of S as the total mass gets smaller, as we can see in the plot. Therefore, for optimally oriented, equal mass binaries as the ones we are considering here, the memory SNR can give a relevant contribution to the overall SNR especially for systems with aligned positive spins.

Finally, we compute the SNR that we would obtain for an optimally oriented ($\theta=\pi/2$) GW150914-like event. The SNR for this detection was of 24 \cite{LIGOScientific:2016aoc}. Taking into account that such value of SNR would correspond to the dominant modes at this inclination, we calculate the SNR that the $(2,0)$ mode computed with our waveform model, would produce in the Advanced LIGO detector. If we only consider the oscillatory component of this mode, we obtain a value of SNR = 1.11, whereas if we consider also the memory contribution, this value is instead SNR = 1.21. This is far from the threshold for detection (SNR = 8) and we see that for this combination of parameters ($q\approx1.22$, $\chi_1\approx0.32$, $\chi_2\approx-0.44$, $M\approx65M_{\odot}$, $D_L\approx410$ Mpc, $z\approx0.88$), the increase of the SNR when considering the memory is almost negligible. Our conclusion is consistent with the findings in the recent paper \cite{Chen:2024ieh} which investigates the detectability of the memory effect with LIGO O4 sensitivity and future-generation detectors such as Cosmic Explorer (CE) \cite{CE} and LISA \cite{LISA}. In any case, as we have seen in the results presented above, for a more optimal combination of parameters that favor the SNR of the $(2,0)$ mode, the contribution of the memory could raise it above the detection threshold. So far, studies concerning single events from the GWTC-1 and GWTC-2 catalogs have not found any evidence of memory, as expected \cite{Hubner:2019sly,Ebersold:2020zah,Hubner:2021amk}. While detecting memory with a single event seems unlikely with the current generation of detectors, stacking multiple ($\sim$1000) events can help extracting the memory information from a population of merging black holes \cite{Lasky:2016knh,Hubner:2019sly,Boersma:2020gxx}. Moreover, the detection threshold is expected to be surpassed routinely when CE and ET become operational \cite{Goncharov:2023woe,Grant:2022bla,Chen:2024ieh}. 

\section{Conclusions}
\label{sec:conclusions}


Our main result is the development of a model that describes the full $(2,0)$ spherical harmonic, encompassing both the memory contribution and the ringdown oscillations. Our model follows the strategy of modeling other spherical harmonic modes within the ``phenomenological waveforms program'',
e.g. in the {\tt IMRPhenomXHM} \cite{Garcia-Quiros:2020qpx} and {\tt IMRPhenomTHM} \cite{Estelles:2020twz} models, and is based on piecewise closed form expressions,  which can be evaluated efficiently. Our model is calibrated against NR simulations \cite{SXS:catalog} that contain the QNM ringdown used for the oscillatory contribution and the memory waveforms we obtain from calculations based on the BMS balance laws.

We have compared the results when one uses only the dominant modes as input for the BMS law calculation, and when also including higher modes, leading us to the conclusion to include at least the $(3,\pm2)$ modes as they give the most important contribution after the $(2,\pm2)$.

We have tested the model's accuracy by comparing it to the corresponding SXS simulations, and two preexisting models \cite{Cao_2016,Yoo:2023spi}. In relation to the detectability of this mode with Advanced LIGO detector, it has been observed that introducing memory increases the SNR for certain parameters of the system. We checked that, although the memory contribution is considerable in some cases, the largest contribution arises from the oscillatory component of the $(2,0)$ mode.

The model of this spherical harmonic has been implemented as an extension of the phenomenological model {\tt IMRPhenomTHM} in LALSuite. We plan to make the code publicly available in the future. Regarding the performance, the time needed to produce this mode is of the same order as for the other modes implemented in {\tt IMRPhenomTHM}. For instance, we test it with an equal-mass, equal-spin system, with $f_{\text{min}}=10\;$Hz, $f_{\text{max}}=2096\;$Hz and a sampling interval of $1/4096$ s, in a range of total masses between 10 and 500$M_{\odot}$. A single evaluation of the (2,0) mode for the lowest mass takes around 1.3 seconds, and for the highest mass, 0.0038 seconds. We check that, within a small variation, this evaluation time is the same as for the other modes in the model. Note that the {\tt IMRPhenomTHM} code has not been designed for small masses/long waveforms, where  {\tt IMRPhenomXHM} is more accurate and efficient, e.g. due to the use of the multibanding acceleration method \cite{Garcia-Quiros:2020qlt}.
We expect that future versions of {\tt IMRPhenomTHM} will permit much faster evaluations of waveforms which contain long inspiral portions.

Many further improvements can be made to our model. First, it will be beneficial to increase its accuracy, especially for systems with high positive spins. Further work will also be required to extend the applicability of the model to higher mass ratios and to match with the EMRI limit.  More importantly, however, the model needs to be generalized for BBH systems with misaligned spins and eccentric orbits. 

\newpage
\section*{Acknowledgments}
We thank Denis Pollney for providing some NR waveforms \cite{Pollney:2010hs} for the (2,0) mode which were used for comparison during the work. We also thank Scott A. Hughes, Anuj Apte, Gaurav Khanna and Halston Lim for providing the EMRI waveforms used in this project \cite{Apte:2019txp,Lim:2019xrb}. We thank Jorge Valencia G\'omez and Rodrigo Tenorio for discussions about numerical Fourier transforms.

Maria Rossell\'o-Sastre is supported by the Spanish Ministry of Universities via an FPU doctoral grant (No. FPU21/05009).

This work was supported by the Universitat de les Illes Balears (UIB); the Spanish Agencia Estatal de Investigaci\'on grants No. PID2022-138626NB-I00, No. RED2022-134204-E, No. RED2022-134411-T, funded by MICIU/AEI/10.13039/501100011033, by the ESF+ and ERDF/EU; the MICIU with funding from the European Union NextGenerationEU/PRTR (PRTR-C17.I1); the Comunitat Aut\`onoma de les Illes Balears through the Direcci\'o General de Recerca, Innovaci\'o I Transformaci\'o Digital with funds from the Tourist Stay Tax Law (PDR2020/11 - ITS2017-006), the Conselleria d’Economia, Hisenda i Innovaci\'o grants No. SINCO2022/18146 and No. SINCO2022/6719, cofinanced by the European Union and FEDER Operational Program 2021-2027 of the Balearic Islands.

\appendix

\begin{widetext}
\section{Calculation of the memory effect including higher modes }
\label{appendix:BMS_HM}

We decompose the strain as the sum of the modes we are considering
\begin{equation}
\begin{split}
    \dot{h}(t,\theta,\phi)&\approx \dot{h}_{2,2}(t)\;^{-2}Y_{2,2}(\theta,\phi)+\dot{h}_{2,-2}(t)\;^{-2}Y_{2,-2}(\theta,\phi)+\dot{h}_{2,1}(t)\;^{-2}Y_{2,1}(\theta,\phi)+\dot{h}_{2,-1}(t)\;^{-2}Y_{2,-1}(\theta,\phi)+\dot{h}_{3,2}(t)\;^{-2}Y_{3,2}(\theta,\phi)\\
    &+\dot{h}_{3,-2}(t)\;^{-2}Y_{3,-2}(\theta,\phi)+\dot{h}_{3,3}(t)\;^{-2}Y_{3,3}(\theta,\phi)+\dot{h}_{3,-3}(t)\;^{-2}Y_{3,-3}(\theta,\phi)+\dot{h}_{4,4}(t)\;^{-2}Y_{4,4}(\theta,\phi)+\dot{h}_{4,-4}(t)\;^{-2}Y_{4,-4}(\theta,\phi).
\end{split}
\end{equation}
After decomposing the strain into the real and imaginary parts and computing the absolute value squared, we project the result of $|\dot{h}|^2$ onto the basis of $s=0$ spherical harmonics.  In the following equations we see that the functions $^0Y_{\ell,m}$ with $\ell$ odd give no contribution.  For the $^0Y_{\ell,m}$ with $\ell$ even we expect to get the same contributions as before from the $(2,\pm2)$ mode and additional contributions from the other modes.

For the $^0Y_{2,0}$,  we get a contribution from all the modes except from the $(3,\pm3)$, the mode $(3,\pm2)$ appears in a mixed term together with the $(2,\pm2)$,
\begin{equation}
\label{proj20}
    \int_{\phi=0}^{2\pi}\int_{\theta=0}^{\pi}\; ^0Y_{2,0}(\theta)\; \left|\dot{h}\right|^2 \sin\theta\; d\theta d\phi=\frac{2}{7}\sqrt{\frac{5}{\pi}}\left|\dot{h}_{2,2}\right|^2-\frac{1}{7}\sqrt{\frac{5}{\pi}}\left|\dot{h}_{2,1}\right|^2+\frac{5}{7}\sqrt{\frac{7}{\pi}}\left(\dot{h}_{2,2}^{\text{Re}}\dot{h}_{3,2}^{\text{Re}}+\dot{h}_{2,2}^{\text{Im}}\dot{h}_{3,2}^{\text{Im}}\right)-\frac{8}{11\sqrt{5\pi}}\left|\dot{h}_{4,4}\right|^2.
\end{equation}
The projection onto the $^0Y_{3,0}$ gives 
\begin{equation}
\label{proj30}
    \int_{\phi=0}^{2\pi}\int_{\theta=0}^{\pi}\; ^0Y_{3,0}(\theta)\; \left|\dot{h}\right|^2 \sin\theta\; d\theta d\phi=0.
\end{equation}
The $(4,0)$ mode has a contribution from all the considered modes
\begin{equation}
\label{proj40}
\begin{split}
    \int_{\phi=0}^{2\pi}\int_{\theta=0}^{\pi}\; ^0Y_{4,0}(\theta)\; \left|\dot{h}\right|^2 \sin\theta\; d\theta d\phi=&\frac{1}{42\sqrt{\pi}}\left|\dot{h}_{2,2}\right|^2-\frac{2}{21\sqrt{\pi}}\left|\dot{h}_{2,1}\right|^2-\frac{7}{22\sqrt{\pi}}\left|\dot{h}_{3,3}\right|^2+\frac{2}{21}\sqrt{\frac{35}{\pi}}\left(\dot{h}_{2,2}^{\text{Re}}\dot{h}_{3,2}^{\text{Re}}+\dot{h}_{2,2}^{\text{Im}}\dot{h}_{3,2}^{\text{Im}}\right)\\
    &+\frac{49}{66\sqrt{\pi}}\left|\dot{h}_{3,2}\right|^2-\frac{3}{13\sqrt{\pi}}\left|\dot{h}_{4,4}\right|^2.
\end{split}
\end{equation}
The $^0Y_{5,0}$ gives a zero contribution
\begin{equation}
\label{proj50}
    \int_{\phi=0}^{2\pi}\int_{\theta=0}^{\pi}\; ^0Y_{5,0}(\theta)\; \left|\dot{h}\right|^2 \sin\theta\; d\theta d\phi=0.
\end{equation}
The $^0Y_{6,0}$ only has contributions from the two highest modes
\begin{equation}
\label{proj60}
    \int_{\phi=0}^{2\pi}\int_{\theta=0}^{\pi}\; ^0Y_{6,0}(\theta)\; \left|\dot{h}\right|^2 \sin\theta\; d\theta d\phi=\frac{3}{11\sqrt{13\pi}}\left|\dot{h}_{3,2}\right|^2-\frac{1}{22\sqrt{13\pi}}\left|\dot{h}_{3,3}\right|^2+\frac{2}{5\sqrt{13\pi}}|\dot{h}_{4,4}|^2.
\end{equation}
As expected,  the $^0Y_{7,0}$ gives no contribution
\begin{equation}
\label{proj70}
    \int_{\phi=0}^{2\pi}\int_{\theta=0}^{\pi}\; ^0Y_{7,0}(\theta)\; \left|\dot{h}\right|^2 \sin\theta\; d\theta d\phi=0.
\end{equation}
And,  finally,  the $^0Y_{8,0}$ only has a contribution from the $(4,\pm 4)$ mode
\begin{equation}
\label{proj80}
    \int_{\phi=0}^{2\pi}\int_{\theta=0}^{\pi}\; ^0Y_{8,0}(\theta)\; \left|\dot{h}\right|^2 \sin\theta\; d\theta d\phi=\frac{14}{715\sqrt{17\pi}}\left|\dot{h}_{4,4}\right|^2.
\end{equation}
Then,  we can express $|\dot{h}|^2$ as a combination of the $s=0$ spherical harmonics with $\ell=2,4,6,8$ as follows
\begin{equation}
\begin{split}
    \left|\dot{h}\right|^2\approx& \left|\dot{h}_{2,2}\right|^2\left[\frac{2}{7}\sqrt{\frac{5}{\pi}}\;^0Y_{2,0}(\theta)+\frac{1}{42\sqrt{\pi}}\;^0Y_{4,0}(\theta)\right]+\left|\dot{h}_{2,1}\right|^2\left[-\frac{1}{7}\sqrt{\frac{5}{\pi}}\;^0Y_{2,0}(\theta)-\frac{2}{21\sqrt{\pi}}\;^0Y_{4,0}(\theta)\right]\\&+\left(\dot{h}_{2,2}^{\text{Re}}\dot{h}_{3,2}^{\text{Re}}+\dot{h}_{2,2}^{\text{Im}}\dot{h}_{3,2}^{\text{Im}}\right)\left[\frac{5}{7}\sqrt{\frac{7}{\pi}}\;^0Y_{2,0}(\theta)+\frac{2}{21}\sqrt{\frac{35}{\pi}}\;^0Y_{4,0}(\theta)\right]\\
    &+\left|\dot{h}_{3,2}\right|^2\left[-\frac{49}{66\sqrt{\pi}}\;^0Y_{4,0}(\theta)-\frac{3}{11\sqrt{13\pi}}\;^0Y_{6,0}(\theta)\right]
    +\left|\dot{h}_{3,3}\right|^2\left[-\frac{7}{22\sqrt{\pi}}\;^0Y_{4,0}(\theta)-\frac{1}{22\sqrt{13\pi}}\;^0Y_{6,0}(\theta)\right]\\
    &+\left|\dot{h}_{4,4}\right|^2\left[-\frac{8}{11\sqrt{5\pi}}\;^0Y_{2,0}(\theta)-\frac{3}{13\sqrt{\pi}}\;^0Y_{4,0}(\theta)+\frac{2}{5\sqrt{13\pi}}\;^0Y_{6,0}(\theta)+\frac{14}{715\sqrt{17\pi}}\;^0Y_{8,0}(\theta)\right].
\end{split}
\end{equation}
To finish the calculation of $J_{\varepsilon}$,  we need to apply the operators $\bar{\eth}^2$ and $\mathfrak{D}^{-1}$ to the previous expression,
\begin{equation}
\label{Jehm}
\begin{split}
    \frac{1}{8}\bar{\eth}^2\mathfrak{D}^{-1} \left|\dot{h}\right|^2\approx& \left|\dot{h}_{2,2}\right|^2\left[\frac{1}{7}\sqrt{\frac{5}{6\pi}}\;^{-2}Y_{2,0}(\theta)+\frac{1}{252\sqrt{10\pi}}\;^{-2}Y_{4,0}(\theta)\right]+\left|\dot{h}_{2,1}\right|^2\left[-\frac{1}{14}\sqrt{\frac{5}{6\pi}}\;^{-2}Y_{2,0}(\theta)-\frac{1}{63\sqrt{10\pi}}\;^{-2}Y_{4,0}(\theta)\right]\\&+\left(\dot{h}_{2,2}^{\text{Re}}\dot{h}_{3,2}^{\text{Re}}+\dot{h}_{2,2}^{\text{Im}}\dot{h}_{3,2}^{\text{Im}}\right)\left[\frac{5}{2\sqrt{42\pi}}\;^0Y_{2,0}(\theta)+\frac{1}{9\sqrt{14\pi}}\;^0Y_{4,0}(\theta)\right]\\
    &+\left|\dot{h}_{3,2}\right|^2\left[\frac{49}{396\sqrt{10\pi}}\;^0Y_{4,0}(\theta)-\frac{1}{44}\sqrt{\frac{3}{455\pi}}\;^0Y_{6,0}(\theta)\right]+\left|\dot{h}_{3,3}\right|^2\left[-\frac{7}{132\sqrt{10\pi}}\;^{-2}Y_{4,0}(\theta)-\frac{1}{88\sqrt{1365\pi}}\;^{-2}Y_{6,0}(\theta)\right]\\
    &+\left|\dot{h}_{4,4}\right|^2\left[-\frac{2}{11}\sqrt{\frac{2}{15\pi}}\;^{-2}Y_{2,0}(\theta)-\frac{1}{130\sqrt{2\pi}}\;^{-2}Y_{4,0}(\theta)+\frac{1}{10\sqrt{1365\pi}}\;^{-2}Y_{6,0}(\theta)+\frac{1}{4290}\sqrt{\frac{7}{85\pi}}\;^{-2}Y_{8,0}(\theta)\right].
\end{split}
\end{equation}
The $s=-2$ spin-weighted spherical harmonics for $\ell=6$ and $\ell=8$ are given by
\begin{equation}
\label{Y60}
    ^{-2}Y_{6,0}(\theta)=\frac{1}{512} \sqrt{\frac{1365}{\pi }} \sin ^2\theta\left[60 \cos (2 \theta )+33 \cos (4 \theta )+35\right],
\end{equation}
\begin{equation}
\label{Y80}
\begin{split}
    &^{-2}Y_{8,0}(\theta)=\frac{3}{16384}\sqrt{\frac{595}{\pi}}\left[64 \cos (2 \theta )+44 \cos (4 \theta )-143 \cos (8 \theta )+35\right].
\end{split}
\end{equation}
We replace (\ref{Y20}),  (\ref{Y40}),  (\ref{Y60}) and (\ref{Y80}) in (\ref{Jehm}) and we obtain the explicit dependence on $\theta$.  As we have done with the dominant modes,  we define as $F_{\ell,\pm m}(\theta)$ each of the terms that depend on $\theta$ multiplying the $|\dot{h}_{\ell,m}|^2$.  The explicit expressions for these multiplicative factors are the following:

\begin{equation}
\label{F22}
    F_{2,\pm2}(\theta)=\frac{\sin ^2\theta \left(17+\cos ^2\theta\right)}{192 \pi },
\end{equation} 
as before,
\begin{equation}
\label{F21}
    F_{2,\pm1}(\theta)=-\frac{\sin ^2\theta \left(2+\cos ^2\theta\right)}{48 \pi },
\end{equation} 
\begin{equation}
     F_{3,\pm2}(\theta)=\frac{\sin ^2\theta (548 \cos (2 \theta )+27 \cos (4 \theta )+385)}{6144 \pi },
\end{equation}
\begin{equation}
    F_{2,\pm2;3,\pm2}(\theta)=\frac{\sqrt{35} \sin ^2\theta\left(2 + \cos ^2\theta \right)}{48 \pi },
\end{equation}
\begin{equation}
\label{F33}
    F_{3,\pm3}(\theta)=-\frac{\sin ^2\theta (148 \cos (2 \theta )+3 (\cos (4 \theta )+35))}{4096 \pi }
\end{equation}
and
\begin{equation}
\label{F44}
    F_{4,\pm4}(\theta)=\frac{\sin ^2\theta \left(3614 \cos (4 \theta )+91 \cos (6 \theta )+\left(6485-2688 \sqrt{5}\right) \cos (2 \theta )-1920 \sqrt{5}-20430\right)}{532480 \pi }.
\end{equation}

 The expression for the memory in this case is given by replacing all these factors into Eq. (\ref{integrand}).

\newpage
\section{Qualitative comparison of the models and the SXS simulations}
\label{appendix:compmodel}

We present in Fig. \ref{fig:plotsmodels} the qualitative comparison of the waveforms obtained with the models and the SXS simulations for the same combinations of parameters as for the mismatch studies presented in Fig. \ref{fig:MmodelsSXS}. 

\begin{center}
\begin{figure}[htp]
\includegraphics[width=0.75\textwidth]{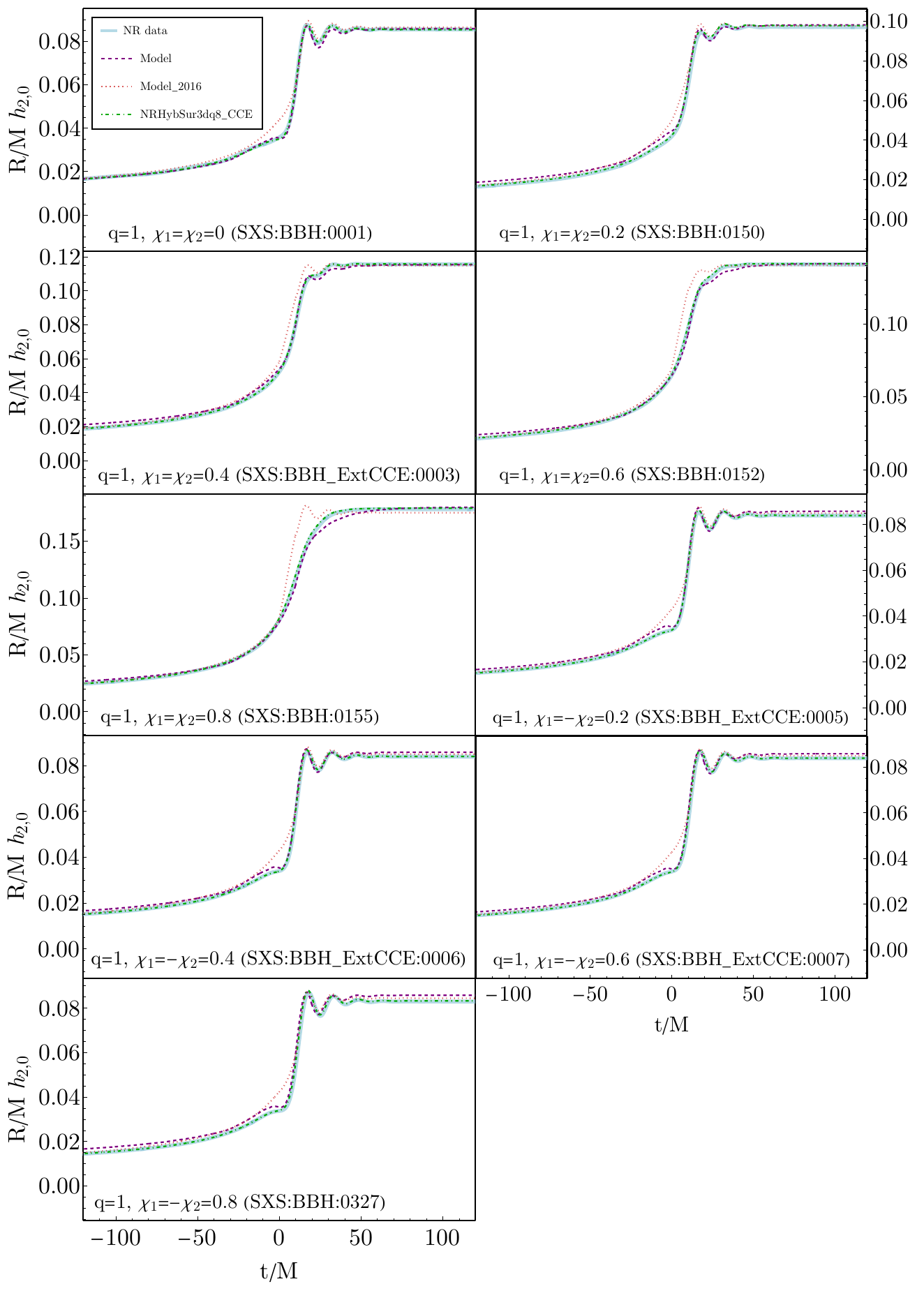}
    \caption{Comparison of the waveforms at late times obtained with our model, the model in \cite{Cao_2016} and the model in \cite{Mitman:2022kwt} with the corresponding SXS simulations from \cite{SXS:catalog, SXS:ExtCCE} for the parameters stated in the plots. For the visual comparison the model waveforms are shifted to match the off-set at zero frequency of the numerical waveforms, therefore $h_{2,0}(f=0)\neq0$. The legend in the upper left plot is common to all of them. 
    \label{fig:plotsmodels}
    }
\end{figure}
\end{center}
\end{widetext}




\let\c\Originalcdefinition %
\let\d\Originalddefinition %
\let\i\Originalidefinition

\clearpage
\bibliography{final_bibliography}

\end{document}